\documentclass[aps,prfluids,groupedaddress,onecolumn,10pt, superscriptaddress, notitlepage]{revtex4}

\usepackage{lipsum}
\usepackage{amsmath}
\usepackage{amssymb}
\usepackage{amsthm}
\usepackage{bbm}
\usepackage{graphicx}
\usepackage{color}
\usepackage{upgreek}
\usepackage[linkcolor = blue, citecolor = blue, urlcolor = blue, colorlinks = true]{hyperref}
\usepackage{amssymb}
\usepackage{bm}
\usepackage[capitalise]{cleveref}
\usepackage{textcomp}
\usepackage{hyperref}
\usepackage[usenames,dvipsnames]{xcolor}
\usepackage[normalem]{ulem}
\usepackage{wasysym}
\usepackage{mathtools}
\usepackage{subeqnarray}

 \renewcommand{\vec}[1]{\mathbf{#1}}
\newcommand{\vect}[1]{\boldsymbol{\mathbf{#1}}}

\renewcommand{\vec}[1]{\mathbf{#1}}
\renewcommand{\imath}[0]{\mathsf{i}}

\hypersetup{colorlinks=true, linkcolor=blue, urlcolor=blue, citecolor=blue}

\definecolor{ABpurple}{RGB}{128, 0, 128}
\definecolor{ABred}{RGB}{255, 0, 0}
\definecolor{ABgreen}{RGB}{0, 255, 0}
\definecolor{ABbrown}{RGB}{128, 64, 0}
\definecolor{ABblue}{RGB}{0, 0, 255}

\newcommand*{\defeq}{\stackrel{\text{def}}{=}}

\begin{document}
\title{A note about convected time derivatives for flows of complex fluids}


\author{Howard A. Stone}
\email{hastone@princeton.edu}
\affiliation{Department of Mechanical and Aerospace Engineering, Princeton University, New Jersey 08544, USA}
\author{Michael J. Shelley}
\email{mshelley@flatironinstitute.org}
\affiliation{Courant Institute of Mathematical Sciences, New York University, New York, NY 10012, USA}
\affiliation{Center for Computational Biology, Flatiron Institute, Simons Foundation, New York, NY 10010, USA}

\author{Evgeniy Boyko}
\email{evgboyko@technion.ac.il}
\affiliation{Department of Mechanical and Aerospace Engineering, Princeton University, New Jersey 08544, USA}
\affiliation{Faculty of Mechanical Engineering, Technion--Israel Institute of Technology, Haifa, 3200003 Israel}

\date{\today}

\begin{abstract}

We present a direct derivation of the typical time derivatives used in a continuum description of 
complex fluid flows, harnessing the principles of the kinematics of line elements.  The evolution of the microstructural conformation tensor in a flow and the physical interpretation of different derivatives then follow naturally.



\end{abstract}
\maketitle


\section{Introduction}

In the study of the flow of viscoelastic or complex fluids it is necessary to utilize constitutive descriptions that capture the local deformations, e.g., reorientations, stretching, compression, that give rise to the stress in a material element following the flow~\cite{bird1987dynamics1,bird1987dynamics2,larson1988constitutive}. Although perhaps simple to state in words, as in the previous sentence, the mathematical description is complex as the stress and material deformation, the latter  expressed via the velocity gradients at the position of the fluid element, are second-rank tensorial quantities. 

Most presentations of the mathematical descriptions begin by describing the need for material objectivity of a tensorial property  when translating, rotating, and deforming with the local flow, then give (messy) details of covariant and contravariant characterizations of vectors and tensors. Thus, different constitutive descriptions, utilizing different time derivatives of tensorial quantities, e.g., known as upper- or lower-convected derivatives, are obtained. These steps lead to the forms of the stress versus rate-of-strain relation known as the Oldroyd-B or Oldroyd-A constitutive description~\cite{oldroyd1950formulation,bird1987dynamics1,larson1988constitutive}. It is reasonable to assume that most readers do not find such mathematical, Cartesian tensor presentations helpful for forming a physical picture of the underlying dynamics.

Alternatively, a physical picture of a dilute polymer solution can be gained by considering a dumbbell model~\cite{kuhn1934gestalt,bird1987dynamics2,larson1988constitutive,Intro_C_F}. It is well known that a dumbbell model, written moving with flow, provides a description of the stress versus rate-of-strain relation, expressed in terms of the conformation tensor, that yields the upper-convected time derivative and the Oldroyd-B constitutive description~\cite{lumley1971applicability,bird1987dynamics2,hinch2021oldroyd,beris2021continuum,datta2021perspectives,edwards2023oldroyd}. The dumbbell model also allows various physically plausible modifications~\cite{hinch1974mechanical,hinch1977mechanical,de1974coil,bird1980polymer,fuller1980flow,fuller1981effects,phan1984study,dunlap1987dilute,chilcott1988creeping,harrison1998dynamics,remmelgas1999computational}.

Here, we provide a direct derivation of the typical time derivatives using the idea of kinematics of line elements. The physical interpretation, and even some of the steps, have overlap with the common derivation of the dumbbell description of the state of stress, but we are not otherwise aware of the style of presentation we provide here (see also \cite[][Sec. 12.4]{VenerusOttinger}). We believe that the physical picture associated with this derivation, focusing as it does on the material line kinematics, which should be familiar from a standard graduate mechanics course, is more transparent than existing alternatives as to the origin of terms in the time derivative associated with the material deformation.  It is then straightforward to discuss other uses of the time derivatives in the development of constitutive descriptions.


\section{Kinematics of a line element}\label{Sec2}
\subsection{The material derivative of a line element}
\begin{figure}[t]
 \centerline{\includegraphics[scale=0.7]{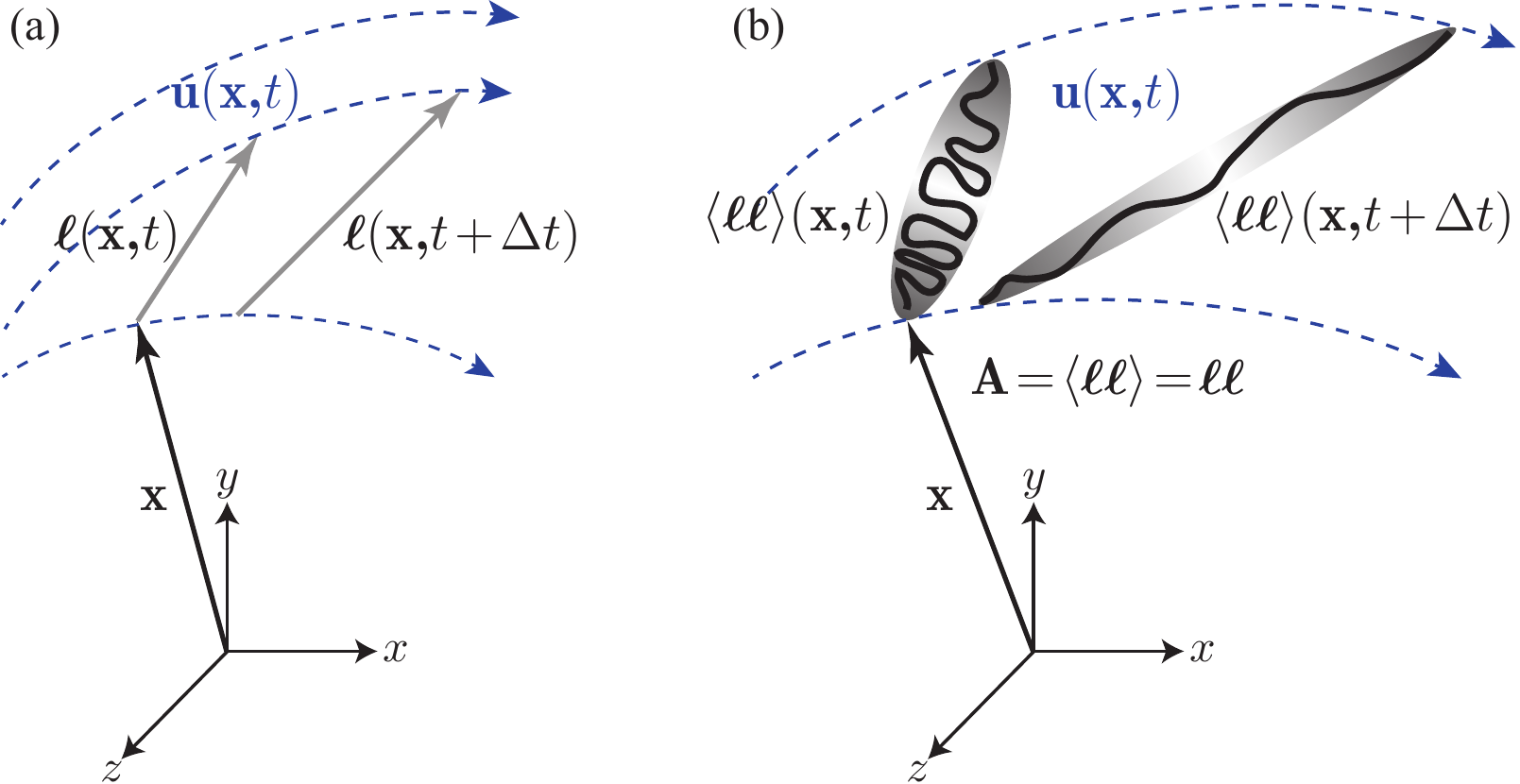}}\caption{Kinematics of a material line element $\boldsymbol{\ell}$ and conformation tensor $\vec{A}=\left \langle \boldsymbol{\ell}\boldsymbol{\ell}\right \rangle$.}
\label{F1}
\end{figure}

Consider an Eulerian velocity field $\vec{u}\left(\vec{x}, t\right )$ at position $\vec{x}$ and time $t$. Recall the definition of the material derivative,
\begin{equation}
    \frac{D}{Dt} =\frac{\partial}{\partial t}+\vec{u}\cdot\boldsymbol{\nabla}.\label{MatlDerivative0}
\end{equation}
It is shown in fluid dynamics textbooks that discuss kinematics~(see, e.g.,~\cite[][pp. 131--132]{batchelor2000introduction}) that an infinitesimal line element $\boldsymbol{\ell}\left(\vec{x}, t\right )$ moving with the local fluid velocity is reoriented and changed in length according to
\begin{equation}
    \frac{D\boldsymbol{\ell}}{Dt} = \boldsymbol{\ell}\cdot\boldsymbol{\nabla}\vec{u}, \label{MatlDerivative1}
\end{equation} as illustrated in Fig.~\ref{F1}(a). 
Next, we introduce a time derivative representing the motion of a line element in the flow,
\begin{equation}
  \stackrel{*}{\boldsymbol{\ell}} \defeq \frac{D\boldsymbol{\ell}}{Dt}  -\boldsymbol{\ell} \cdot\boldsymbol{\nabla}\vec{u}.\label{DerivativeLineElementFollowingFlow}
\end{equation}It follows that $\stackrel{*}{\boldsymbol{\ell}} = \boldsymbol{0}$ 
means that a line element translates with the local fluid velocity,
and rotates and stretches (or compresses) according to the local velocity gradient~\cite{hinch2021oldroyd}. In Sec. 
\ref{MatDerTensor}, we apply this idea to second-rank tensors. 

 \subsection{Lagrangian trajectories of a line element and a theorem of Cauchy on invariance}\label{SubSec2Lag}

Before proceeding, we link the above ideas to a Lagrangian description, which makes clear that time derivatives are really expressing invariances associated with following a vectorial or tensorial quantity along the trajectory of a fluid particle. These features have a direct connection to a theorem of Cauchy, which is most simply understood using a Lagrangian description. 
In this subsection, we discuss the ideas for a vector field, and in subsection \ref{InvarianceConformationTensor} we show how the same ideas apply to the conformation tensor (see also \cite{snoeijer2020relationship}).
  
 To describe motion, we follow Lagrangian trajectories, $\vec{x}=\vec{x}(t;\vec{X})$, where $\vec{x}(0;\vec{X})=\vec{X}$, with the label $\vec{X}$ indicating the initial position. 
   A differential displacement $d\vec{X}$  is mapped to a  differential displacement $d\vec{x}$ according to
\begin{equation}
  d\vec{x}=d\vec{X}\cdot \vec{F},
\end{equation}where $\vec{F}(t;\vec{X}) = \nabla_{\vec{X}}\vec{x}$ is the deformation gradient tensor; in index notation we write, $F_{ij} = \frac{\partial x_j}{\partial X_i}$. 
Taking the material time derivative, we write in a Lagrangian description 
$  \frac{D\vec{F}}{D t} = \nabla_{\vec{X}}\frac{\partial\vec{x}}{\partial t}=\nabla_{\vec{X}}\vec{u}^L$, where the Lagrangian velocity is $\vec{u}^L (t;\vec{X})=
\vec{u}\left (\vec{x}, t\right )$. Using the chain rule, we can involve the Eulerian representation as $\nabla_{\vec{X}}\vec{u}=\vec{F}\cdot\vect{\nabla}\vec{u} $. Therefore, the material derivative of $\vec{F}$ at fixed $\vec{X}$ (as in Eq.~(\ref{MatlDerivative0})) is
\begin{equation}
    \left.\frac{\partial \vec{F}}{\partial t}\right|_{\vec{X}}=\frac{D\vec{F}}{D t} =\vec{F}\cdot\vect{\nabla}\vec{u}.
    \label{DerivativeofF}
\end{equation}
Since by definition $\vec{F}^{-1}\cdot \vec{F}=\vec{I}$, where $\vec{I}$ is the identity tensor, then differentiating with respect to time means $
    \vec{F}^{-1}\cdot \frac{\partial\vec{F}}{\partial t}=-\frac{\partial\vec{F}^{-1}}{\partial t}\cdot\vec{F}$, which are two identities (at fixed $\vec{X}$) that will prove useful. The reader should note that in some presentations, Eq.~(\ref{DerivativeofF}), because of a notational choice, appears with the transpose~(see, e.g.,~\cite{Intro_C_F}).

Given these geometric preliminaries, for the (infinitesimal) line element transport equation (\ref{MatlDerivative1}) we use the identity tensor in the form $\vec{F}^{-1}\cdot \vec{F}=\vec{I}$ and write ($\vect{\ell}(\vec{x},t)=\vect{\ell}_{L}(t;\vec{X})$),
\begin{equation}
   \frac{D\vect{\ell}}{Dt}=\left.\frac{\partial\vect{\ell}_{L}}{\partial t}\right|_{\vec{X}}= \vect{\ell}\cdot\vec{F}^{-1}\cdot\underbrace{\vec{F}\cdot\vect{\nabla}\vec{u}}_{\frac{\partial\vec{F}}{\partial t}}=
    \vect{\ell}_{L}\cdot\vec{F}^{-1}\cdot\frac{\partial\vec{F}}{\partial t}=-\vect{\ell}_{L}\cdot\frac{\partial\vec{F}^{-1}}{\partial t}\cdot\vec{F}.
\end{equation}Taking the inner product on the right with
$\vec{F}^{-1}$, we then have for fixed $\vec{X}$,
\begin{equation}
    \frac{\partial \vect{\ell}_{L}}{ \partial t}\cdot\vec{F}^{-1}+\vect{\ell}_{L}\cdot\frac{\partial\vec{F}^{-1}}{\partial t}=\vec{0}\quad\hbox{or}\quad
    \frac{\partial}{ \partial t}\left (\vect{\ell}_{L}\cdot\vec{F}^{-1}\right )=\vec{0}.
\end{equation}We conclude that $\vect{\ell}_{L}\cdot\vec{F}^{-1}=\hbox{constant}$ following the fluid motion, i.e., it is an invariant. Since initially $\vec{F}(0)=\vec{I}$, then 
$ \vect{\ell}_{L}\cdot\vec{F}^{-1}=\vect{\ell}_{L}(0)$, the initial vectorial line element. As $\vec{F}$ is a function of time that is in principle known, once a  velocity field is determined, we conclude that the line element, as a Lagrangian object,  evolves in time according to 
\begin{equation}
    \vect{\ell}_{L}(t;\vec{X})= \vect{\ell}_{L}(0)\cdot \vec{F}(t;\vec{X}).
\label{MappingLineElement11}
\end{equation}This result is known as Cauchy's theorem and yields how any initial line element is changed in magnitude and orientation by the time-dependent deformation tensor $\vec{F}$. The short derivation presented here may be helpful to readers to see the origin of this idea. Below we indicate the extension of these ideas to second-rank tensors, such as the conformation tensor and the convected time derivatives.

\section{The material derivative of a second-rank tensor}\label{MatDerTensor}

\subsection{Conformation tensor}

Consider an end-to-end vector $\vect{r}$, whose magnitude at
equilibrium is $r_{\rm eq}$. We define the dimensionless conformation tensor $\vec{A}\defeq \left \langle \vect{r}\vect{r}\right \rangle/r_{\rm eq}^2 \defeq \left \langle \boldsymbol{\ell}\boldsymbol{\ell}\right \rangle$, where $\left\langle \cdot\right \rangle$ denotes the ensemble average and the dyadic product $\boldsymbol{\ell \ell}$ characterizes the state of deformation of a given microstructure, as sketched in Fig.~\ref{F1}(b); for example, $\vec{A}=\vec{I}$ represents an undeformed isotropic  (equilibrium) state.
For simplicity of the notation, we drop 
$\left\langle \cdot\right \rangle $ in the calculation that follows.

Proceeding formally to take the material derivative, and applying the product rule for differentiation, we obtain
\begin{equation}
  \frac{D}{Dt}\vec{A}= \frac{D}{Dt} \left ( \boldsymbol{\ell}\boldsymbol{\ell}\right )= \frac{D\boldsymbol{\ell}}{Dt}\boldsymbol{\ell}+ \boldsymbol{\ell}\frac{D\boldsymbol{\ell}}{Dt}.\label{MaterialDerivativeDyad1}
  \end{equation}Using Eq. (\ref{MatlDerivative1}) and the properties of the dyadic product, we find
\begin{equation}
  \frac{D}{Dt}\vec{A}=\frac{D}{Dt} \left ( \boldsymbol{\ell}\boldsymbol{\ell}\right )= \boldsymbol{\ell} \cdot\left (\boldsymbol{\nabla}\vec{u}\right )\boldsymbol{\ell}+ \boldsymbol{\ell}\boldsymbol{\ell} \cdot\left (\boldsymbol{\nabla}\vec{u}\right )= \left (\boldsymbol{\nabla}\vec{u}\right )^T\cdot \vec{A}+\vec{ A}\cdot \boldsymbol{\nabla}\boldsymbol{u}.
  \end{equation} This identity is also recognized in  Ref.~\cite[][Sec. 12.4]{VenerusOttinger}.
Therefore, we introduce a time derivative for the change of the tensor (dyadic product) deforming in a fluid flow:
  \begin{equation}
\stackrel{\triangledown}{\vec{A}}\defeq\frac{D}{Dt}\vec{A}-\left (\boldsymbol{\nabla}\vec{u}\right )^T\cdot \vec{A}-\vec{A}\cdot \boldsymbol{\nabla}\vec{u}. \label{Oldroyd-B time derivative}
  \end{equation}
  The derivative $\stackrel{\triangledown}{\vec{A}}$
  is known as the upper-convected or  contravariant  time derivative~\cite{bird1987dynamics1}. We  conclude that $\stackrel{\triangledown}{\vec{A}}={\bf 0}$ means that the conformation tensor deforms exactly as does the corresponding line elements in a flow. In contrast, 
   $\stackrel{\triangledown}{\vec{A}}\ne{\bf 0}$ means that $\vec{A}$ does not deform comparable to the line elements in the flow.

   \subsection{Arbitrary dyadic products and second-rank tensors}

As a further remark, we note that the kinematic approach here is valid for any dyadic product.
Since an arbitrary second-rank tensor, such as the stress tensor, can be written as a linear combination of three dyadic products~\cite[][Secs. 61--63]{LBrand}, then it follows that the derivation of the time derivatives discussed above also applies to an arbitrary second-rank tensor. For example, if we define the dyadic product ${\vec{B}}={\vec{a}}{\vec{b}}$, where $\vec{a}$ and $\vec{b}$ are vectors, then taking the material derivative and treating $\vec{a}$ and $\vec{b}$ as line elements (see Eq.~(\ref{MatlDerivative1})), we obtain
\begin{equation}
  \frac{D}{Dt}\vec{B}= \frac{D}{Dt} \left ( \vec{a}\vec{b}\right )= \frac{D\vec{a}}{Dt}\vec{b}+ \vec{a}\frac{D\vec{b}}{Dt}= \vec{a} \cdot\left (\boldsymbol{\nabla}\vec{u}\right )\vec{b}+ \vec{a}\vec{b} \cdot\left (\boldsymbol{\nabla}\vec{u}\right )= \left (\boldsymbol{\nabla}\vec{u}\right )^T\cdot \vec{B}+\vec{ B}\cdot \boldsymbol{\nabla}\vec{u},
  \end{equation}which is the form of the upper-convected derivative.

\subsection{Invariance associated with convected time derivatives of the conformation tensor}\label{InvarianceConformationTensor}

Recall Sec.~\ref{Sec2} on the kinematics of line elements, where the (Eulerian) time derivative $
  \stackrel{*}{\boldsymbol{\ell}} =\boldsymbol{0}$ following the flow (Eq.~(\ref{DerivativeLineElementFollowingFlow})) implies the invariant solution $\vect{\ell}_{L}\cdot\vec{F}^{-1}=\hbox{constant}$, from which the  time variation of $\vect{\ell}_{L} (t; \vec{X})$ can be determined for a given flow or $\vec{F}$ (see Eq.~(\ref{MappingLineElement11})). Following similar steps, one can take a Lagrangian view and integrate the equation 
$\stackrel{\triangledown}{\vec{A}}=
\frac{D\vec{A}}{Dt}-\left(\boldsymbol{\nabla}\vec{u}\right)^{T}\cdot\vec{A}-\vec{A}\cdot\boldsymbol{\nabla}\vec{u}=\boldsymbol{0}$. These ideas are known in the literature and the notation from Sec.~\ref{Sec2} provides a compact way to proceed (see Supplementary Material). In particular, using a Lagrangian notation with $\vec{A}(\vec{x},t)=\vec{A}_{L}(t; \vec{X})$ one can show that $\stackrel{\triangledown}{\vec{A}}=\boldsymbol{0}$ leads to
\begin{equation}
\frac{\partial}{\partial t} 
\left ( \vec{F}^{-T}\cdot\vec{A}_{L}\cdot\vec{F}^{-1}
\right )=\boldsymbol{0}.
\end{equation}It follows that $\vec{F}^{-T}\cdot\vec{A}_{L}\cdot\vec{F}^{-1}=\hbox{constant}$,
 following the fluid motion, i.e., it is an invariant. Using the initial data, we obtain $\vec{F}^{-T}\cdot\vec{A}_{L}\cdot\vec{F}^{-1}=\vec{A}_{L}(0;\vec{X})$ since $\vec{F}(0)=\vec{I}$, which means $\vec{A}_{L}(t; \vec{X})=\vec{F}^{T}\cdot\vec{A}_{L}(0;\vec{X})\cdot\vec{F}$. In other words, the initial data and $\vec{F}$ determine the evolution of the conformation tensor in this case.

  \section{The Jeffery equation for a particle in a linear flow}

  \subsection{A line element that lags the flow}
  
  A slender extensible filament or rod with orientation and length $\boldsymbol \ell$, not necessarily infinitesimal, in a flow with vorticity $\boldsymbol{\omega}=\boldsymbol{\nabla}\wedge\vec{u}$ and rate-of-strain tensor $\vec{E}=(1/2)(\boldsymbol{\nabla}\vec{u}+(\boldsymbol{\nabla}\vec{u})^{T})$ reorients approximately according to~(see, e.g.,~\cite{ericksen1960transversely,gordon1971bead,gordon1972anisotropic})
  \begin{equation}
      \frac{D\boldsymbol{\ell}}{Dt} = \frac{1}{2}\boldsymbol{\omega}\wedge\boldsymbol{\ell}+\beta \boldsymbol{\ell}\cdot\vec{E},
      \label{JefferyEquation1}
  \end{equation}where $\beta$ is a constant.
   A rigid rod is unable to change length so Eq. (\ref{JefferyEquation1}) is then modified to preserve length by adding an extra term $-\beta\boldsymbol{\ell}\cdot\vec{E}\cdot\boldsymbol{\ell}\boldsymbol{\ell}$ to the right-hand side, resulting in the Jeffrey equation~\cite{jeffery1922motion} (see 
 Sec.~\ref{Sec5}(vii)).
   Recalling  the vorticity tensor $\boldsymbol{\Omega}=(1/2)(\boldsymbol{\nabla}\vec{u}-(\boldsymbol{\nabla}\vec{u})^{T})$, where $\boldsymbol{\nabla}\vec{u}=\vec{E}+\boldsymbol{\Omega}$ and $(1/2)\boldsymbol{\omega}\wedge\boldsymbol{\ell}=\boldsymbol{\ell}\cdot\boldsymbol{\Omega}$, Eq.~(\ref{JefferyEquation1}) can be written as
   \begin{equation}
      \frac{D\boldsymbol{\ell}}{Dt} =\boldsymbol{\ell}\cdot\boldsymbol{\nabla}\vec{u}+\left (\beta -1\right )\boldsymbol{\ell}\cdot \vec{E},
      \label{JefferyEquation2}
  \end{equation}where the last term indicates that the straining motion produces non-affine motion of the filament, relative to following the motion and deformation of the fluid. Therefore, a filament behaves like a line element and follows the flow for $\beta=1$, and lags the deformation for $\beta<1$, as arises below.  

  \subsection{A surface element}

In the same way that the divergence of the velocity vector expresses the fractional rate of change of volume ($V)$ of an infinitesimal fluid element in a flow (so is zero for an incompressible flow), and a line element $\boldsymbol{\ell}$ changes it length and orientation according to Eq.~(\ref{MatlDerivative1}), one can consider an infinitesimal (vectorial) surface element $\vec{S}$, i.e., $V=\boldsymbol{\ell}\cdot\vec{S}$. It is an exercise to show that for an incompressible flow, $\boldsymbol{\nabla}\cdot\vec{u}=0$, then (see, e.g.,~\cite[][pp. 131--132]{batchelor2000introduction} and \cite{stone2017fundamentals,eggers2023rheology})
\begin{equation}
    \frac{D\vec{S}}{Dt} = -\left( \nabla \vec{u}\right )\cdot \vec{S} = \frac{1}{2} \boldsymbol{\omega}\wedge\vec{S}- \vec{S}\cdot\vec{E}.
      \label{JefferyEquation3Surface}
\end{equation}Hence, we see that an infinitesimal surface element satisfies Eq.~(\ref{JefferyEquation1}) with $\beta=-1$.

  \section{Non-affine  motion of the conformation tensor}\label{Sec5}

The derivation of Eq. (\ref{Oldroyd-B time derivative}), and the  steps that now follow, are the main contributions of this note focusing on the meaning of time derivatives for the kinematics of fluids with deformable microstructure. 
  Substituting Eq. (\ref{JefferyEquation2}) into Eq. (\ref{MaterialDerivativeDyad1}), we obtain 
  \begin{equation}
  \frac{D}{Dt} \vec{A}=\frac{D}{Dt} \left ( \boldsymbol{\ell}\boldsymbol{\ell}\right )= \left (\boldsymbol{\nabla}\vec{u}\right )^T\cdot \vec{A}+\vec{A}\cdot \boldsymbol{\nabla}\vec{u}+\left (\beta-1\right )\left 
  (\vec{ A} \cdot\vec{E}+\vec{E}\cdot\vec{A}\right),
  \label{GeneralDerivative}
  \end{equation}or, using the definition of $\stackrel{\triangledown}{\vec{A}}$,
  \begin{equation}
\stackrel{\triangledown}{\vec{A}}= \frac{D}{Dt} \vec{A}- \left (\boldsymbol{\nabla}\vec{u}\right )^T\cdot \vec{A}-\vec{A}\cdot \boldsymbol{\nabla}\vec{u}=\left (\beta-1\right )\left 
  (\vec{A} \cdot\vec{E}+\vec{E}\cdot\vec{A}\right).
  \label{GeneralDerivative2}
  \end{equation}The right-hand side of Eq. (\ref{GeneralDerivative2}) yields non-affine motion that changes the response from purely following the deformation of the fluid, i.e., we no longer have $\stackrel{\triangledown}{\vec{A}}={\bf 0}$. We next make 
  several observations.
  \begin{enumerate}
  \item[(i)] $\beta=1$: We observe that $\beta=1$ gives back the case of a line element following the flow, i.e., the upper-convected  derivative vanishes, i.e., $\stackrel{\triangledown}{\vec{A}}={\bf 0}$. 
  
      \item[(ii)] $\beta = 0$: Using the relation $\boldsymbol{\Omega}=\boldsymbol{\nabla}\vec{u}-\vec{E}$, we obtain
      \begin{equation}
\stackrel{\circ}{\vec{A}}\defeq\frac{D}{Dt}\vec{A}-\boldsymbol{\Omega}^T\cdot \vec{A}-\vec{A}\cdot \boldsymbol{\Omega}=\boldsymbol{0}.
  \end{equation}$\stackrel{\circ}{\vec{A}}$ 
  is referred to as the (Jaumann) co-rotational derivative~\cite{zaremba1903remarques,jaumann1911geschlossenes}, representing the rate of change of a fluid element advected
with the flow and rotating with the local vorticity, e.g., see Eq.~(\ref{JefferyEquation1}).

      \item[(iii)] $\beta = -1$: 
    In this case, using the definition for the rate-of-strain tensor $\vec{E}=(1/2)(\boldsymbol{\nabla}\vec{u}+(\boldsymbol{\nabla}\vec{u})^{T})$, Eq. (\ref{GeneralDerivative}) becomes
     \begin{equation}
 \frac{D}{Dt} \vec{A}= \left (\boldsymbol{\nabla}\vec{u}\right )^T\cdot \vec{A}+\vec{A}\cdot \boldsymbol{\nabla}\vec{u}-2\left 
  (\vec{A} \cdot\vec{E}+\vec{E}\cdot\vec{A}\right)=  -\left (\boldsymbol{\nabla}\vec{u}\right )\cdot \vec{A}-  \vec{A}\cdot\left ( \boldsymbol{\nabla}\vec{u}\right )^T.
  \label{GeneralDerivativewithbetaminus1}
  \end{equation}Similar to Eq.~(\ref{Oldroyd-B time derivative}), we can introduce the lower-convected or covariant time derivative (see, e.g., \cite{bird1987dynamics1,larson1988constitutive,hinch2021oldroyd}):
  \begin{equation}
\overset{\vartriangle}{\vec{A}}\defeq \frac{D}{Dt} \vec{A}+
 \left (\boldsymbol{\nabla}\vec{u}\right )\cdot \vec{A}+  \vec{A}\cdot\left ( \boldsymbol{\nabla}\vec{u}\right )^T.
  \label{LowerConvectedDerivative}
  \end{equation}
  \item[(iv)] $\beta=\mathrm{const}$: For a constant value of $\beta$, using Eq.~(\ref{GeneralDerivative}), we can introduce the Gordon$-$Schowalter convected time derivative $\overset{\square}{\vec{A}}$ (see, e.g., \cite{gordon1972anisotropic,larson1988constitutive,hinch2021oldroyd}),
  \begin{equation}
\stackrel{\square}{\vec{A}}\defeq \frac{D}{Dt} \vec{A}- \left (\boldsymbol{\nabla}\vec{u}\right )^T\cdot \vec{A}-\vec{A}\cdot \boldsymbol{\nabla}\vec{u}-\left (\beta-1\right )\left 
  (\vec{A} \cdot\vec{E}+\vec{E}\cdot\vec{A}\right).
  \label{GordonSchowalterDerivative}
  \end{equation}
  \item[(v)] 
  We note that, in practice, $\beta$, appearing in Eq.  (\ref{GordonSchowalterDerivative}), can be a function of the conformation tensor $\vec{A}$. In fact, for a dumbbell model, \citet{hinch1977mechanical} and \citet{phan1984study} pointed out that $\beta$ depends on the trace of $\vec{A}$ via
  \begin{equation}
\beta=\frac{\mathrm{tr}\vec{A}-a}{\mathrm{tr}\vec{A}+b},
  \label{beta}
  \end{equation}
where $\mathrm{tr}\vec{A}$ represents the square of the dumbbell extension and $a$ and $b$ are constants (see also \cite{rallison1988we,hinch2021oldroyd}). 

\item[(vi)] An interesting property of the conformation tensor $\vec{A}$ that satisfies $\stackrel{\triangledown}{\vec{A}}={\bf 0}$ (upper-convected  derivative) is that its inverse $\vec{A}^{-1}$ satisfies $\left(\vec{A}^{-1}\right)^{\vartriangle}={\bf 0}$ (lower-convected derivative). This result can be readily shown by left and right multiplying $\stackrel{\triangledown}{\vec{A}}={\bf 0}$ by $\vec{A}^{-1}$ and using the identity $\frac{\partial \vec{A}^{-1}}{\partial t}=-\vec{A}^{-1}\cdot\frac{\partial \vec{A}} {\partial t}\cdot\vec{A}^{-1}$. Similarly, $\stackrel{\circ}{\vec{A}}={\bf 0}$ (co-rotational derivative) implies that $\left(\vec{A}^{-1}\right)^{\circ}={\bf 0}$. In fact, this symmetric property is more general and can be extended to the Gordon$-$Schowalter convected time derivative $\overset{\square}{\vec{A}}$, that if $\overset{\square}{\vec{A}}_{\beta}={\bf 0}$ for some $\beta$, then $\left(\vec{A}^{-1}\right)_{-\beta}^{\square}={\bf 0}$ with $-\beta$.


\item[(vii)] 
    For a rigid inextensible rod, we modify Eq. (\ref{JefferyEquation1}) to preserve the length of the rod by including an additional term $-\beta\boldsymbol{\ell}\cdot\vec{E}\cdot\boldsymbol{\ell}\boldsymbol{\ell}$ on the right-hand side. Thus, we have the Jeffrey equation for the motion of a rigid ellipsoid\textcolor{blue}{~\cite{jeffery1922motion}}, 
\begin{equation}
\frac{D\boldsymbol{\ell}}{Dt}=\frac{1}{2}\boldsymbol{\omega}\wedge\boldsymbol{\ell}+\beta\left(\boldsymbol{\ell}\cdot\vec{E}-\boldsymbol{\ell}\cdot\vec{E}\cdot\boldsymbol{\ell}\boldsymbol{\ell}\right),\label{JefferyEquatioInextensible1}
\end{equation}
which also can be written as
\begin{equation}
\frac{D\boldsymbol{\ell}}{Dt}=\boldsymbol{\ell}\cdot\boldsymbol{\nabla}\vec{u}+(\beta-1)\boldsymbol{\ell}\cdot\vec{E}-\beta\boldsymbol{\ell}\cdot\vec{E}\cdot\boldsymbol{\ell}\boldsymbol{\ell}.\label{JefferyEquatioInextensible2}
\end{equation}
 Here, the parameter $\beta$ is $\beta=(r^2-1)/(r^2+1)$, where $r$ is the length-to-radius aspect ratio of an axisymmetric ellipsoid~\cite{jeffery1922motion}.
Substituting Eq.~(\ref{JefferyEquatioInextensible1}) into Eq.~(\ref{MaterialDerivativeDyad1}), we obtain the time-rate-of-change of the corresponding conformation tensor,
  \begin{equation}
  \frac{D}{Dt} \vec{A}=\frac{D}{Dt} \left ( \boldsymbol{\ell}\boldsymbol{\ell}\right )= \left(\boldsymbol{\nabla}\vec{u}\right)^{T}\cdot\vec{A}+\vec{A}\cdot\boldsymbol{\nabla}\vec{u}+(\beta-1)\left(\vec{A}\cdot\vec{E}+\vec{E}\cdot\vec{A}\right)-2\beta\vec{A}\cdot\vec{E}\cdot\vec{A}.
  \label{Derivative-Doi-type theories1}
  \end{equation}
  Using the definition of $\stackrel{\triangledown}{\vec{A}}$ yields,
  \begin{equation}
\stackrel{\triangledown}{\vec{A}}= \frac{D}{Dt} \vec{A}- \left (\boldsymbol{\nabla}\vec{u}\right )^T\cdot \vec{A}-\vec{A}\cdot \boldsymbol{\nabla}\vec{u}=\left (\beta-1\right )\left 
  (\vec{ A} \cdot\vec{E}+\vec{E}\cdot\vec{A}\right)-2\beta\vec{A}\cdot\vec{E}\cdot\vec{A}.
  \label{Derivative-Doi-type theories2}
  \end{equation}Again, we see how $\beta$ causes the conformation tensor to evolve differently than a ``passive" structure simply following the fluid flow, $\stackrel{\triangledown}{\vec{A}}\ne {\bf 0}$. 
As noted by \citet{hinch2021oldroyd}, fibers, long thin rods, and elongated particles with $r\gg1$ behave like line elements (of zero thickness) for which $\beta\approx1$, so that the upper-convected derivative is appropriate. On the other hand, flattened particles with $r\ll1$ behave like area elements, so that $\beta\approx -1$, and thus, the lower-convected derivative is appropriate.
  

\item[(viii)] As a particular case of the time derivative in Eq.~(\ref{Derivative-Doi-type theories2}), consider the case with $\beta=1$. We refer to this time derivative as the constrained upper-convected time derivative, given as
\begin{equation}
\stackrel{\triangledown}{\vec{A}}+2\vec{A}\cdot\vec{E}\cdot\vec{A}= \frac{D}{Dt} \vec{A}- \left (\boldsymbol{\nabla}\vec{u}\right )^T\cdot \vec{A}-\vec{A}\cdot \boldsymbol{\nabla}\vec{u}+2\vec{A}\cdot\vec{E}\cdot\vec{A}=\boldsymbol{0}.
  \label{Derivative-Doi-type theories3}
  \end{equation}
This time derivative arises, for example, in the so-called quadratic closure for the Doi-Onsager rod theory as shown in~\citet{weady2022thermodynamically} and in the sharply aligned case of the Doi-Onsager rod theory \cite{gao2017analytical}.
It is possible to extend the ideas introduced above (Sections \ref{Sec2}  and \ref{InvarianceConformationTensor}) for identifying invariances following the fluid motion to the special case of $\stackrel{\triangledown}{\vec{A}}\ne{\bf 0}$ when $\stackrel{\triangledown}{\vec{A}}=-2\vec{A}\cdot\vec{E}\cdot\vec{A}$.
  
For example, using a Lagrangian notation with $\vec{A}(\vec{x},t)=\vec{A}_{L}(t; \vec{X})$ one can show (see Supplementary Material) that Eq. (\ref{Derivative-Doi-type theories3}) leads to 
 \begin{equation}
\frac{\partial}{\partial t} 
\left ( \vec{F}^{-T}\cdot\vec{A}_{L}\cdot\vec{F}^{-1}\right )=-\left ( \vec{F}^{-T}\cdot\vec{A}_{L}\cdot\vec{F}^{-1}\right )\cdot\frac{\partial}{\partial t}\left ( \vec{F}\cdot\vec{F}^{T}\right )\cdot\left ( \vec{F}^{-T}\cdot\vec{A}_{L}\cdot\vec{F}^{-1}
\right ).
\end{equation}
or  \begin{equation}
\frac{\partial \vec{G}_{L}}{\partial t}=-\vec{G}_{L}\cdot\frac{\partial}{\partial t}\left ( \vec{F}\cdot\vec{F}^{T}\right )\cdot\vec{G}_{L}\qquad\mathrm{with}\qquad \vec{G}_{L}=\vec{F}^{-T}\cdot\vec{A}_{L}\cdot\vec{F}^{-1}. \label{G_L}
\end{equation}
Under the assumption that $\vec{G}_{L}$ (in fact, $\vec{A}_{L}$) is \emph{invertible} and using the identities $\vec{G}_{L}^{-1}=\vec{F}\cdot\vec{A}_{L}^{-1}\cdot\vec{F}^{T}$ and $\frac{\partial \vec{G}_{L}}{\partial t}=-\vec{G}_{L}\cdot\frac{\partial \vec{G}_{L}^{-1}}{\partial t}\cdot\vec{G}_{L}$, from Eq. (\ref{G_L}) it follows that
\begin{equation}
\frac{\partial \vec{G}_{L}^{-1}}{\partial t}=\frac{\partial}{\partial t}\left ( \vec{F}\cdot\vec{F}^{T}\right ) \qquad\mathrm{or}\qquad 
\frac{\partial}{\partial t} \left (\vec{G}_{L}^{-1}-\vec{F}\cdot\vec{F}^{T}\right )=\frac{\partial}{\partial t} \left (\vec{F}\cdot\vec{A}_{L}^{-1}\cdot\vec{F}^{T}-\vec{F}\cdot\vec{F}^{T}\right )=\boldsymbol{0}.
\end{equation}
It follows that $\vec{F}\cdot\vec{A}_{L}^{-1}\cdot\vec{F}^{T}-\vec{F}\cdot\vec{F}^{T}=\hbox{constant}$,
 following the fluid motion, i.e., it is an invariant. 
Using the initial data, we obtain $\vec{F}\cdot\vec{A}_{L}^{-1}\cdot\vec{F}^{T}-\vec{F}\cdot\vec{F}^{T}=\vec{A}_{L}^{-1}(0;\vec{X})-\vec{I}$ since $\vec{F}(0)=\vec{I}$, which means $\vec{A}_{L}^{-1}(t; \vec{X})=\vec{I}-\vec{F}^{-1}\cdot\vec{F}^{-T}
+\vec{F}^{-1}\cdot\vec{A}_{L}^{-1}(0;\vec{X})\cdot\vec{F}^{-T}$. Note that if $\vec{A}_{L}(0;\vec{X})=\vec{I}$, then $\vec{A}_{L}=\vec{I}$, which is a steady-state solution of Eq.~(\ref{Derivative-Doi-type theories3}).

By analogy to Eq. (\ref{G_L}) for the constrained upper-convected time derivative, one can show (see Supplementary Material) for the constrained lower-convected time derivative, $\overset{\vartriangle}{\vec{A}}=2\vec{A}\cdot\vec{E}\cdot\vec{A}$, that 
\begin{equation}
\frac{\partial}{\partial t} \left (\vec{H}_{L}^{-1}-\vec{F}^{-T}\cdot\vec{F}^{-1}\right )=\frac{\partial}{\partial t} \left (\vec{F}^{-T}\cdot\vec{A}_{L}^{-1}\cdot\vec{F}^{-1}-\vec{F}^{-T}\cdot\vec{F}^{-1}\right )=\boldsymbol{0}\qquad\mathrm{with}\qquad \vec{H}_{L}=\vec{F}\cdot\vec{A}_{L}\cdot\vec{F}^{T}. \label{H_L}
\end{equation}
Thus, $\vec{F}^{-T}\cdot\vec{A}_{L}^{-1}\cdot\vec{F}^{-1}-\vec{F}^{-T}\cdot\vec{F}^{-1}$ is an invariant following the motion.

For the constrained upper-/lower-convected derivatives, we have conserved quantities $\vec{G}_{L}=\vec{F}^{-T}\cdot\vec{A}_{L}\cdot\vec{F}^{-1}$ and $\vec{H}_{L}=\vec{F}\cdot\vec{A}_{L}\cdot\vec{F}^{T}$, respectively, whose invariant structure is essentially through inverses of each other.
The co-rotational derivative arises from adding these two transport derivatives together, thus suggesting that finding an invariance for the co-rotational time derivative is challenging, if possible at all.




  \end{enumerate}

\section{Connecting conformation and its change to stress}\label{Connecting}
    

  \subsection{The continuity and Cauchy momentum equations with additional microstructural stresses}
    
Although the focus of this note is on the time derivatives, it is natural to now ask about the physical effects that exert forces on suspended filaments, which cause additional changes to these material derivatives and contribute to stress in the fluid. Describing the incompressible hydrodynamics of a complex fluid generally requires incorporating two new equations for the conformation tensor in addition to the continuity and Cauchy momentum equations. The first equation indicates how changes in the conformation give rise to stress in the fluid, whereas the second equation connects the evolution of the conformation to the gradients in the velocity field.  As a specific example, consider the case of a viscoelastic dilute polymer solution, whose  fluid motion is governed by the continuity equation and Cauchy momentum equations,
\begin{equation}
\boldsymbol{\nabla\cdot} \vec{u}=0\qquad\hbox{and}\qquad\rho\frac{D\vec{u}}{Dt}=\boldsymbol{\nabla\cdot}\boldsymbol{\sigma},\label{Continuity Momentum}
\end{equation}
where $\rho$ is fluid density and $\boldsymbol{\sigma}$ is the stress
tensor given by
\begin{equation}
\boldsymbol{\sigma}=-p\vec{I}+2\eta_{s}\vec{E}+\boldsymbol{\tau}_{p}.\label{Stress tensor viscoelastic}
\end{equation}
The first term on the right-hand side of Eq. (\ref{Stress tensor viscoelastic})
is the pressure contribution, the second term is the viscous stress
contribution of Newtonian solvent with a constant viscosity $\eta_{s}$,
and the last term, $\boldsymbol{\tau}_{p}$, is the polymer contribution
to the stress tensor. The latter contribution arises due to the response
of polymer chains to fluid flow so that their microstructure deviates
from the equilibrium state and deforms, $\vec{A}\neq\vec{I}$.
This deformation results in a polymer contribution
to the stress tensor, which, in the simplest Hookean case, can be expressed
via $\vec{A}$ as (see, e.g., \cite{snoeijer2020relationship,hinch2021oldroyd,beris2021continuum,datta2021perspectives})
\begin{equation}
\boldsymbol{\tau}_{p}=G(\vec{A})(\vec{A}-\vec{I}),\label{tau_p}
\end{equation}
where $G(\vec{A})$ is the elastic modulus, which may be a function
of $\vec{A}$. Equations (\ref{Continuity Momentum})$-$(\ref{tau_p})
 show the coupling between the fluid velocity and deformation of microstructure.~Understanding this two-way coupling also requires determining the evolution equation for the conformation changes that describes how polymer chains deform in the flow.

If we treat polymer chains like
line elements transported and freely deformed by the flow, it implies
$\stackrel{\triangledown}{\vec{A}}= {\bf 0}$, according to Eq. (\ref{Oldroyd-B time derivative}). However, polymer chains have elastic features with an inherent restoring mechanism opposing stretching/compression in the flow and causing relaxation to an equilibrium (coiled) state in the absence of flow.
 In
general, this relaxation process may be characterized by several time
scales; we denote by $\lambda$ the longest relaxation time of
the polymers. Thus, for affine motion ($\beta=1$), accounting for
the relaxation to an undeformed state on a time scale $\lambda$, the
evolution equation for $\vec{A}$ can be written as 
\begin{equation}
\stackrel{\triangledown}{\vec{A}}=-\frac{1}{\lambda(\vec{A})}(\vec{A}-\vec{I}),\label{OB}
\end{equation}
which, for a constant $\lambda$, is known as the Oldroyd-B constitutive
equation \cite{hinch2021oldroyd,beris2021continuum}. We note that $\lambda(\vec{A})$ may be a function
of $\vec{A}$, as was first elucidated by \citet{de1974coil} and \citet{hinch1974mechanical}.

From Eq. (\ref{OB}), it follows that when the observation time $t_{\rm obs}$
is much shorter than the relaxation time $\lambda$, $t_{ \rm obs}\ll\lambda$,
$\stackrel{\triangledown}{\vec{A}}\approx{\bf 0}$ and thus,
polymer chains behave like line elements that are transported and
deformed by the flow without relaxing. We note that the ratio $\lambda/t_{\rm obs}$ is known as the Deborah number, $De\defeq\lambda/t_{\rm obs}$~\cite{bird1987dynamics1,bird1987dynamics2}, which follows from Eq.~(\ref{OB}). In many cases, the observation time $t_{\rm obs}$ is based on the convective time scale $L_c/u_c$, where $L_c$ is the characteristic length scale and $u_c$ is the characteristic velocity, so that the Deborah number is $De=\lambda u_c/L_c$.
Furthermore, as many flows are characterized  by a shear rate, $\dot{\gamma}$, Eq.~(\ref{OB}) also naturally introduces the Weissenberg number, $Wi=\lambda\dot{\gamma}$.
While, for brevity, we have considered only the Oldroyd-B model, the approach illustrated here allows
various physically plausible modifications similar to the modeling of dumbbells~\cite{hinch2021oldroyd,hinch1974mechanical,hinch1977mechanical,de1974coil,bird1980polymer,fuller1980flow,fuller1981effects,phan1984study,dunlap1987dilute,chilcott1988creeping,harrison1998dynamics,remmelgas1999computational}.  
Finally, we refer the reader to the recent work by \citet{eggers2023rheology} on flat elastic particles in a Newtonian fluid that, by analogy with dumbbell models, were modeled as three beads connected by nonlinear springs. In this work, a lower-convected time derivative naturally arises as part of the constitutive equation.

\subsection{Lagrangian integration of the Oldroyd-B constitutive equation}

The ideas introduced above (Sections \ref{Sec2}  and \ref{InvarianceConformationTensor}) for identifying invariances following the fluid motion can also be applied to the evolution equation for the conformation tensor when elastic stresses are included. In this case, the conformation tensor evolves according to the deformation tensor $\vec{F}$ and the time constant $\lambda_0$. Again, the detailed steps can be found in the literature (see, e.g.,~\cite[][pp. 431--432]{bird1987dynamics1}), \cite{hohenegger2011dynamics,snoeijer2020relationship} and Supplementary Material). 

For example, 
 we can consider the Oldroyd-B constitutive equation (Eq.~(\ref{OB})), $
\stackrel{\triangledown}{\vec{A}}=-\lambda_{0}^{-1}(\vec{A}-\vec{I})$, where $\lambda_{0}$ is a constant relaxation time. Again, we use the notation 
$\vec{A}(\vec{x},t)=\vec{A}_{L}(t; \vec{X})$.
The equation for $\stackrel{\triangledown}{\vec{A}}$ can be rearranged to find an ordinary differential equation for $\vec{F}^{-T}\cdot\vec{A}_{L}\cdot\vec{F}^{-1}$ with a time-dependent forcing, 
\begin{equation}
\frac{\partial}{\partial t} 
\left ( \vec{F}^{-T}\cdot\vec{A}_{L}\cdot\vec{F}^{-1}
\right )+\frac{1}{\lambda_0}\vec{F}^{-T}\cdot\vec{A}_{L}\cdot\vec{F}^{-1}=\frac{1}{\lambda_0}\vec{F}^{-T}\cdot\vec{F}^{-1}.
\end{equation}This equation is solved with initial data (suppressing dependence on $\vec{X}$) 
$\vec{F}^{-T}\cdot\vec{A}_{L}\cdot\vec{F}^{-1}=\vec{A}_{L}(0)$ since $\vec{F}(0)=\vec{I}$. Therefore, integrating, we obtain (see also e.g.~\cite{hohenegger2011dynamics,snoeijer2020relationship}),
\begin{equation}
\vec{A}_{L}(t)=\hbox{e}^{-t/\lambda_0} \vec{F}^{T}(t)\cdot\vec{A}_{L}(0)\cdot\vec{F}(t)+
\frac{1}{\lambda_0}\vec{F}^{T}(t)\cdot\int_0^t\vec{F}^{-T}\left (t^\prime\right )\cdot\vec{F}^{-1}\left (t^\prime\right ) \hbox{e}^{-(t-t^\prime)/\lambda_0}~{\rm d}t^\prime\cdot\vec{F}(t).
\end{equation}Thus, we observe that the flow history matters and the material is always relaxing, relative to the local flow features, according to an exponential in time. The reader is referred to \cite{snoeijer2020relationship} for more discussion and the connection of these ideas to the theory of elastic solids.

\section{Concluding remarks} 

The approach illustrated here for the derivation of the typical time derivatives of the conformation tensor, Eqs. (\ref{Oldroyd-B time derivative}) and (\ref{GeneralDerivative2}), commonly used in the study of the flow of complex fluids, is the main contribution of this note. Our derivation utilizes the idea of the kinematics of line elements, and we believe that the physical picture associated with this derivation may be useful in providing insight and developing intuition when using these time derivatives and the corresponding constitutive equations. For completeness, in Sec.~\ref{Connecting}, we briefly discussed the connection of the flow kinematics to the conformation changes that produce stresses. In several places in the discussion, we highlighted how a Lagrangian view of the kinematics allows one to identify invariances following the flow, or, in the case of elastic stresses, how the conformation tensor relaxes in a time-varying way relative to the stretch expected due to the flow alone.
   
\section*{Conflicts of interest}
There are no conflicts to declare.
   
  \begin{acknowledgments}
H.A.S.\ is grateful for partial support of the work by NSF through Princeton University’s Materials Research Science and Engineering Center DMR-2011750. E.B.\ acknowledges the support of the Yad Hanadiv (Rothschild) Foundation, the Zuckerman STEM Leadership Program, and the Lillian Gilbreth Postdoctoral Fellowship from Purdue’s College of Engineering. We thank Jens Eggers, Tannie Liverpool, and Alex Mietke for helpful feedback.
\end{acknowledgments}



\pagebreak
\widetext
\begin{center}
\textbf{\large Supplementary Material for\\
A note about convected time derivatives for flows of complex fluids}
\end{center}
\setcounter{section}{0}
\setcounter{figure}{0}
\setcounter{equation}{0}
\setcounter{figure}{0}
\setcounter{table}{0}
\setcounter{page}{1}
\makeatletter
\renewcommand{\theequation}{S\arabic{equation}}
\renewcommand{\thefigure}{S\arabic{figure}}
\renewcommand{\thesection}{S.\arabic{section}}
\renewcommand{\thesubsection}{\thesection.\arabic{subsection}}
\renewcommand{\thesubsubsection}{\thesubsection.\arabic{subsubsection}}

\section{Invariance associated with convected time derivatives of the conformation tensor}

Recall Sec. II in the manuscript on the kinematics of line elements, where the (Eulerian) time derivative $
  \stackrel{*}{\boldsymbol{\ell}} =\boldsymbol{0}$ following the flow ($  \stackrel{*}{\boldsymbol{\ell}} = \frac{D\boldsymbol{\ell}}{Dt}  -\boldsymbol{\ell} \cdot\boldsymbol{\nabla}\vec{u}$) implies the invariant solution $\vect{\ell}_{L}\cdot\vec{F}^{-1}=\hbox{constant}$, from which the  time variation of $\vect{\ell}_{L} (t; \vec{X})$ can be determined for a given flow or $\vec{F}$ (see Eq.~(\ref{MappingLineElement11})). Following similar steps, one can take a Lagrangian view and integrate the equation 
$\stackrel{\triangledown}{\vec{A}}=
\frac{D\vec{A}}{Dt}-\left(\boldsymbol{\nabla}\vec{u}\right)^{T}\cdot\vec{A}-\vec{A}\cdot\boldsymbol{\nabla}\vec{u}=\boldsymbol{0}$. We use the identities 
$\vec{F}^{-1}\cdot\vec{F}=\vec{I}$,  $\vec{F}^{T}\cdot\vec{F}^{-T}=\vec{I}$, 
$\frac{\partial\vec{F}^{-T}}{\partial t}=-\vec{F}^{-T}\cdot\left(\boldsymbol{\nabla}\vec{u}\right)^{T}
$,  $\frac{\partial\vec{F}}{\partial t}=\vec{F}\cdot\boldsymbol{\nabla}\vec{u}$,
and take a Lagrangian notation with $\vec{A}(\vec{x},t)=\vec{A}_{L}(t; \vec{X})$ to write
\begin{eqnarray}
\stackrel{\triangledown}{\vec{A}}&=&
\frac{\partial\vec{A}_{L}}{\partial t}-\vec{F}^{T}\cdot\vec{F}^{-T}\cdot\left(\boldsymbol{\nabla}\vec{u}\right)^{T}\cdot\vec{A}_{L}-\vec{A}_{L}\cdot\vec{F}^{-1}\cdot\vec{F}\cdot\boldsymbol{\nabla}\vec{u} 
\nonumber \\
&=&
\frac{\partial\vec{A}_{L}}{\partial t}+\vec{F}^{T}\cdot\frac{\partial\vec{F}^{-T}}{\partial t}\cdot\vec{A}_{L}-\vec{A}_{L}\cdot\vec{F}^{-1}\cdot\frac{\partial\vec{F}}{\partial t}=\frac{\partial \vec{A}_{L}}{\partial t}+\vec{F}^{T}\cdot\frac{\partial\vec{F}^{-T}}{\partial t}\cdot\vec{A}_{L}+\vec{A}_{L}\cdot\frac{\partial\vec{F}^{-1}}{\partial t}\cdot\vec{F}.\label{S1}
\end{eqnarray}Right multiplying by $\vec{F}^{-1}$ and left multiplying by $\vec{F}^{-T}$ yields 
\begin{equation}
\vec{F}^{-T}\cdot\stackrel{\triangledown}{\vec{A}}
\cdot\vec{F}^{-1} = \vec{F}^{-T}\cdot
\frac{\partial\vec{A}_{L}}{\partial t}
\cdot\vec{F}^{-1} +\frac{\partial\vec{F}^{-T}}{\partial t}\cdot\vec{A}_{L}\cdot\vec{F}^{-1}+\vec{F}^{-T}\cdot\vec{A}_{L}\cdot\frac{\partial\vec{F}^{-1}}{\partial t}=\frac{\partial}{\partial t} 
\left ( \vec{F}^{-T}\cdot\vec{A}_{L}\cdot\vec{F}^{-1} \right ). \label{S2}
\end{equation}Thus, we conclude that 
$\stackrel{\triangledown}{\vec{A}}=\boldsymbol{0}$ implies 
\begin{equation}
\frac{\partial}{\partial t} 
\left ( \vec{F}^{-T}\cdot\vec{A}_{L}\cdot\vec{F}^{-1}
\right )=\boldsymbol{0}\quad\hbox{or}\quad  \vec{F}^{-T}\cdot\vec{A}_{L}\cdot\vec{F}^{-1}=\hbox{constant},
\end{equation} following the fluid motion, i.e., it is an invariant. Using the initial data, $\vec{F}^{-T}\cdot\vec{A}_{L}\cdot\vec{F}^{-1}=\vec{A}_{L}(0;\vec{X})$ since $\vec{F}(0)=\vec{I}$, which means $\vec{A}_{L}=\vec{F}^{T}\cdot\vec{A}_{L}(0;\vec{X})\cdot\vec{F}$.
For example, if $\vec{A}_{L}(0;\vec{X})=\vec{I}$, i.e., an undeformed isotropic  state, then $\vec{A}_{L}=\vec{F}^{T}\cdot\vec{F}$.

Similarly, one can show for the lower-convected time derivative that $\overset{\vartriangle}{\vec{A}}=\boldsymbol{0}$
implies $\frac{\partial}{\partial t}\left(\vec{F}\cdot\vec{A}_{L}\cdot\vec{F}^{T}\right)=\boldsymbol{0}$. Thus, $\vec{F}\cdot\vec{A}_{L}\cdot\vec{F}^{T}$ is an invariant following the motion.

\section{Invariance associated with the constrained convected time derivative of the conformation tensor}

It is possible to extend the ideas introduced above  for identifying invariances following the fluid motion when $\stackrel{\triangledown}{\vec{A}}=\boldsymbol{0}$ to the case of $\stackrel{\triangledown}{\vec{A}}=-2\vec{A}\cdot\vec{E}\cdot\vec{A}$, corresponding to the constrained upper-convected time derivative.
 
Using a Lagrangian notation with $\vec{A}(\vec{x},t)=\vec{A}_{L}(t; \vec{X})$ and Eq.~(\ref{S1}), $\stackrel{\triangledown}{\vec{A}}=-2\vec{A}\cdot\vec{E}\cdot\vec{A}$ can be expressed as
 \begin{equation}
\stackrel{\triangledown}{\vec{A}}=\frac{\partial \vec{A}_{L}}{\partial t}+\vec{F}^{T}\cdot\frac{\partial\vec{F}^{-T}}{\partial t}\cdot\vec{A}_{L}+\vec{A}_{L}\cdot\frac{\partial\vec{F}^{-1}}{\partial t}\cdot\vec{F}=-2\vec{A}_{L}\cdot\vec{E}\cdot\vec{A}_{L}.\label{S4}
\end{equation}
Right multiplying by $\vec{F}^{-1}$ and left multiplying by $\vec{F}^{-T}$ and using Eq.~(\ref{S2}),
$\vec{F}^{-1}\cdot\vec{F}=\vec{I}$,  and $\vec{F}^{T}\cdot\vec{F}^{-T}=\vec{I}$, Eq.~(\ref{S4})  yields 
\begin{equation}
\frac{\partial}{\partial t} 
\left ( \vec{F}^{-T}\cdot\vec{A}_{L}\cdot\vec{F}^{-1}\right )=-2\left ( \vec{F}^{-T}\cdot\vec{A}_{L}\cdot\vec{F}^{-1}\right )\cdot \left ( \vec{F}\cdot\vec{E}\cdot\vec{F}^{T}\right )\cdot\left ( \vec{F}^{-T}\cdot\vec{A}_{L}\cdot\vec{F}^{-1}
\right). \label{LagrangianTrajectory4}
\end{equation}
Using the identities $\frac{\partial\vec{F}}{\partial t}=\vec{F}\cdot\boldsymbol{\nabla}\vec{u}$ and  $\frac{\partial\vec{F}^{T}}{\partial t}=(\boldsymbol{\nabla}\vec{u})^{T} \cdot \vec{F}^{T}$, the term $2( \vec{F}\cdot\vec{E}\cdot\vec{F}^{T})$ in Eq.~(\ref{LagrangianTrajectory4}) can be expressed as 
\begin{equation}
2\left ( \vec{F}\cdot\vec{E}\cdot\vec{F}^{T}\right )= \vec{F}\cdot\left (\boldsymbol{\nabla}\vec{u}+(\boldsymbol{\nabla}\vec{u})^{T}\right )\cdot\vec{F}^{T}=\frac{\partial\vec{F}}{\partial t} \cdot\vec{F}^{T}+\vec{F}\cdot\frac{\partial\vec{F}^{T}}{\partial t}=\frac{\partial}{\partial t}\left ( \vec{F}\cdot\vec{F}^{T}\right ), 
\end{equation}so that Eq.~(\ref{LagrangianTrajectory4}) takes the form
\begin{equation}
\frac{\partial}{\partial t} 
\left ( \vec{F}^{-T}\cdot\vec{A}_{L}\cdot\vec{F}^{-1}\right )=-\left ( \vec{F}^{-T}\cdot\vec{A}_{L}\cdot\vec{F}^{-1}\right )\cdot\frac{\partial}{\partial t}\left ( \vec{F}\cdot\vec{F}^{T}\right )\cdot\left ( \vec{F}^{-T}\cdot\vec{A}_{L}\cdot\vec{F}^{-1}
\right ).
\end{equation}
or  \begin{equation}
\frac{\partial \vec{G}_{L}}{\partial t}=-\vec{G}_{L}\cdot\frac{\partial}{\partial t}\left ( \vec{F}\cdot\vec{F}^{T}\right )\cdot\vec{G}_{L}\qquad\mathrm{with}\qquad \vec{G}_{L}=\vec{F}^{-T}\cdot\vec{A}_{L}\cdot\vec{F}^{-1}. \label{G_LS}
\end{equation}
Under the assumption that $\vec{G}_{L}$ (in fact, $\vec{A}_{L}$) is \emph{invertible} and using the identities $\vec{G}_{L}^{-1}=\vec{F}\cdot\vec{A}_{L}^{-1}\cdot\vec{F}^{T}$ and $\frac{\partial \vec{G}_{L}}{\partial t}=-\vec{G}_{L}\cdot\frac{\partial \vec{G}_{L}^{-1}}{\partial t}\cdot\vec{G}_{L}$, from Eq. (\ref{G_LS}) it follows that
\begin{equation}
\frac{\partial \vec{G}_{L}^{-1}}{\partial t}=\frac{\partial}{\partial t}\left ( \vec{F}\cdot\vec{F}^{T}\right ) \qquad\mathrm{or}\qquad 
\frac{\partial}{\partial t} \left (\vec{G}_{L}^{-1}-\vec{F}\cdot\vec{F}^{T}\right )=\frac{\partial}{\partial t} \left (\vec{F}\cdot\vec{A}_{L}^{-1}\cdot\vec{F}^{T}-\vec{F}\cdot\vec{F}^{T}\right )=\boldsymbol{0}.
\end{equation}
It follows that $\vec{F}\cdot\vec{A}_{L}^{-1}\cdot\vec{F}^{T}-\vec{F}\cdot\vec{F}^{T}=\hbox{constant}$,
 following the fluid motion, i.e., it is an invariant. 
Using the initial data, we obtain $\vec{F}\cdot\vec{A}_{L}^{-1}\cdot\vec{F}^{T}-\vec{F}\cdot\vec{F}^{T}=\vec{A}_{L}^{-1}(0;\vec{X})-\vec{I}$ since $\vec{F}(0)=\vec{I}$, which means $\vec{A}_{L}^{-1}(t; \vec{X})=\vec{I}-\vec{F}^{-1}\cdot\vec{F}^{-T}
+\vec{F}^{-1}\cdot\vec{A}_{L}^{-1}(0;\vec{X})\cdot\vec{F}^{-T}$. 

Similarly, one can show for the constrained lower-convected time derivative, $\overset{\vartriangle}{\vec{A}}=2\vec{A}\cdot\vec{E}\cdot\vec{A}$, that 
\begin{equation}
\frac{\partial}{\partial t}\left(\vec{F}\cdot\vec{A}_{L}\cdot\vec{F}^{T}\right)=2\left ( \vec{F}\cdot\vec{A}_{L}\cdot\vec{F}^{T}\right )\cdot \left ( \vec{F}^{-T}\cdot\vec{E}\cdot\vec{F}^{-1}\right )\cdot\left ( \vec{F}\cdot\vec{A}_{L}\cdot\vec{F}^{T}
\right). \label{LagrangianTrajectory5}
\end{equation}
Using the identities $\frac{\partial\vec{F}^{-T}}{\partial t}=-\vec{F}^{-T}\cdot\left(\boldsymbol{\nabla}\vec{u}\right)^{T}$ and  $\frac{\partial\vec{F}^{-1}}{\partial t}=-\boldsymbol{\nabla}\vec{u} \cdot \vec{F}^{-1}$, the term $2( \vec{F}^{-T}\cdot\vec{E}\cdot\vec{F}^{-1})$ in Eq.~(\ref{LagrangianTrajectory5}) can be expressed as 
\begin{equation}
2\left ( \vec{F}^{-T}\cdot\vec{E}\cdot\vec{F}^{-1}\right )= \vec{F}^{-T}\cdot\left (\boldsymbol{\nabla}\vec{u}+(\boldsymbol{\nabla}\vec{u})^{T}\right )\cdot\vec{F}^{-1}=-\vec{F}^{-T}\cdot \frac{\partial\vec{F}^{-1}}{\partial t}-\frac{\partial\vec{F}^{-T}}{\partial t} \cdot \vec{F}^{-1}=-\frac{\partial}{\partial t}\left ( \vec{F}^{-T}\cdot\vec{F}^{-1}\right ), 
\end{equation}so that Eq.~(\ref{LagrangianTrajectory5}) takes the form
\begin{equation}
\frac{\partial}{\partial t}\left(\vec{F}\cdot\vec{A}_{L}\cdot\vec{F}^{T}\right)=-\left ( \vec{F}\cdot\vec{A}_{L}\cdot\vec{F}^{T}\right )\cdot \frac{\partial}{\partial t}\left ( \vec{F}^{-T}\cdot\vec{F}^{-1}\right )\cdot\left ( \vec{F}\cdot\vec{A}_{L}\cdot\vec{F}^{T}
\right),
\end{equation}
or  \begin{equation}
\frac{\partial \vec{H}_{L}}{\partial t}=-\vec{H}_{L}\cdot\frac{\partial}{\partial t}\left ( \vec{F}^{-T}\cdot\vec{F}^{-1}\right )\cdot\vec{H}_{L}\qquad\mathrm{with}\qquad \vec{H}_{L}=\vec{F}\cdot\vec{A}_{L}\cdot\vec{F}^{T}. \label{H_LS}
\end{equation}
Again, under the assumption that $\vec{H}_{L}$ (in fact, $\vec{A}_{L}$) is \emph{invertible} and using the identities $\vec{H}_{L}^{-1}=\vec{F}^{-T}\cdot\vec{A}_{L}^{-1}\cdot\vec{F}^{-1}$ and $\frac{\partial \vec{H}_{L}}{\partial t}=-\vec{H}_{L}\cdot\frac{\partial \vec{H}_{L}^{-1}}{\partial t}\cdot\vec{H}_{L}$, from Eq. (\ref{H_LS}) it follows that
\begin{equation}
\frac{\partial \vec{H}_{L}^{-1}}{\partial t}=\frac{\partial}{\partial t}\left (\vec{F}^{-T}\cdot\vec{F}^{-1}\right ) \qquad\mathrm{or}\qquad 
\frac{\partial}{\partial t} \left (\vec{H}_{L}^{-1}-\vec{F}^{-T}\cdot\vec{F}^{-1}\right )=\frac{\partial}{\partial t} \left (\vec{F}^{-T}\cdot\vec{A}_{L}^{-1}\cdot\vec{F}^{-1}-\vec{F}^{-T}\cdot\vec{F}^{-1}\right )=\boldsymbol{0}.
\end{equation}
Thus, $\vec{F}^{-T}\cdot\vec{A}_{L}^{-1}\cdot\vec{F}^{-1}-\vec{F}^{-T}\cdot\vec{F}^{-1}$ is an invariant following the motion.

\section{Lagrangian integration of the Oldroyd-B constitutive equation}

The ideas introduced in Sec. II  and III C for identifying invariances following the fluid motion can also be applied to the evolution equation for the conformation tensor when elastic stresses are included. 
For example, consider the Oldroyd-B constitutive equation (Eq.~(\ref{OB})), $
\stackrel{\triangledown}{\vec{A}}=-\lambda_{0}^{-1}(\vec{A}-\vec{I})$, where $\lambda_{0}$ is a constant relaxation time. In this case, using Eq.~(\ref{S1}) and a Lagrangian notation with $\vec{A}(\vec{x},t)=\vec{A}_{L}(t; \vec{X})$, we have
\begin{equation}
\frac{\partial \vec{A}_{L}}{\partial t}+\vec{F}^{T}\cdot\frac{\partial\vec{F}^{-T}}{\partial t}\cdot\vec{A}_{L}+\vec{A}_{L}\cdot\frac{\partial\vec{F}^{-1}}{\partial t}\cdot\vec{F}=-\frac{1}{\lambda_0}(\vec{A}_{L}-\vec{I}).
\end{equation}Now, recalling the steps just above, right multiplying by $\vec{F}^{-1}$ and left multiplying by $\vec{F}^{-T}$ yields
\begin{equation}
\frac{\partial}{\partial t} 
\left ( \vec{F}^{-T}\cdot\vec{A}_{L}\cdot\vec{F}^{-1}
\right )= -\frac{1}{\lambda_0}\vec{F}^{-T}\cdot(\vec{A}_{L}-\vec{I})\cdot\vec{F}^{-1}.
\end{equation}Thus, we find an ordinary differential equation for 
$\vec{F}^{-T}\cdot\vec{A}_{L}\cdot\vec{F}^{-1}$ with a time-dependent forcing, 
\begin{equation}
\frac{\partial}{\partial t} 
\left ( \vec{F}^{-T}\cdot\vec{A}_{L}\cdot\vec{F}^{-1}
\right )+\frac{1}{\lambda_0}\vec{F}^{-T}\cdot\vec{A}_{L}\cdot\vec{F}^{-1}=\frac{1}{\lambda_0}\vec{F}^{-T}\cdot\vec{F}^{-1}.
\end{equation}This equation is to be solved with initial data (suppressing dependence on $\vec{X}$) 
$\vec{F}^{-T}\cdot\vec{A}_{L}\cdot\vec{F}^{-1}=\vec{A}_{L}(0)$ since $\vec{F}(0)=\vec{I}$. Therefore, integrating, we obtain (see, e.g.~\cite{hohenegger2011dynamics,snoeijer2020relationship}),
\begin{equation}
\vec{F}^{-T}(t)\cdot\vec{A}_{L}(t)\cdot\vec{F}^{-1} (t)=\hbox{e}^{-t/\lambda_0} \vec{A}_{L}(0)+
\frac{1}{\lambda_0}\int_0^t\vec{F}^{-T}\left (t^\prime\right )\cdot\vec{F}^{-1}\left (t^\prime\right ) \hbox{e}^{-(t-t^\prime)/\lambda_0}~{\rm d}t^\prime,
\end{equation}or 
\begin{equation}
\vec{A}_{L}(t)=\hbox{e}^{-t/\lambda_0} \vec{F}^{T}(t)\cdot\vec{A}_{L}(0)\cdot\vec{F}(t)+
\frac{1}{\lambda_0}\vec{F}^{T}(t)\cdot\int_0^t\vec{F}^{-T}\left (t^\prime\right )\cdot\vec{F}^{-1}\left (t^\prime\right ) \hbox{e}^{-(t-t^\prime)/\lambda_0}~{\rm d}t^\prime\cdot\vec{F}(t).
\end{equation}




\begin{thebibliography}{38}%
\makeatletter
\providecommand \@ifxundefined [1]{%
 \@ifx{#1\undefined}
}%
\providecommand \@ifnum [1]{%
 \ifnum #1\expandafter \@firstoftwo
 \else \expandafter \@secondoftwo
 \fi
}%
\providecommand \@ifx [1]{%
 \ifx #1\expandafter \@firstoftwo
 \else \expandafter \@secondoftwo
 \fi
}%
\providecommand \natexlab [1]{#1}%
\providecommand \enquote  [1]{``#1''}%
\providecommand \bibnamefont  [1]{#1}%
\providecommand \bibfnamefont [1]{#1}%
\providecommand \citenamefont [1]{#1}%
\providecommand \href@noop [0]{\@secondoftwo}%
\providecommand \href [0]{\begingroup \@sanitize@url \@href}%
\providecommand \@href[1]{\@@startlink{#1}\@@href}%
\providecommand \@@href[1]{\endgroup#1\@@endlink}%
\providecommand \@sanitize@url [0]{\catcode `\\12\catcode `\$12\catcode
  `\&12\catcode `\#12\catcode `\^12\catcode `\_12\catcode `\%12\relax}%
\providecommand \@@startlink[1]{}%
\providecommand \@@endlink[0]{}%
\providecommand \url  [0]{\begingroup\@sanitize@url \@url }%
\providecommand \@url [1]{\endgroup\@href {#1}{\urlprefix }}%
\providecommand \urlprefix  [0]{URL }%
\providecommand \Eprint [0]{\href }%
\providecommand \doibase [0]{https://doi.org/}%
\providecommand \selectlanguage [0]{\@gobble}%
\providecommand \bibinfo  [0]{\@secondoftwo}%
\providecommand \bibfield  [0]{\@secondoftwo}%
\providecommand \translation [1]{[#1]}%
\providecommand \BibitemOpen [0]{}%
\providecommand \bibitemStop [0]{}%
\providecommand \bibitemNoStop [0]{.\EOS\space}%
\providecommand \EOS [0]{\spacefactor3000\relax}%
\providecommand \BibitemShut  [1]{\csname bibitem#1\endcsname}%
\let\auto@bib@innerbib\@empty
\bibitem [{\citenamefont {Bird}\ \emph
  {et~al.}(1987{\natexlab{a}})\citenamefont {Bird}, \citenamefont {Armstrong},\
  and\ \citenamefont {Hassager}}]{bird1987dynamics1}%
  \BibitemOpen
  \bibfield  {author} {\bibinfo {author} {\bibfnamefont {R.~B.}\ \bibnamefont
  {Bird}}, \bibinfo {author} {\bibfnamefont {R.~C.}\ \bibnamefont
  {Armstrong}},\ and\ \bibinfo {author} {\bibfnamefont {O.}~\bibnamefont
  {Hassager}},\ }\href@noop {} {\emph {\bibinfo {title} {Dynamics of
  {P}olymeric {L}iquids. {V}olume 1: Fluid {M}echanics}}},\ \bibinfo {edition}
  {2nd}\ ed.\ (\bibinfo  {publisher} {John Wiley and Sons, New York},\ \bibinfo
  {year} {1987})\BibitemShut {NoStop}%
\bibitem [{\citenamefont {Bird}\ \emph
  {et~al.}(1987{\natexlab{b}})\citenamefont {Bird}, \citenamefont {Curtiss},
  \citenamefont {Armstrong},\ and\ \citenamefont
  {Hassager}}]{bird1987dynamics2}%
  \BibitemOpen
  \bibfield  {author} {\bibinfo {author} {\bibfnamefont {R.~B.}\ \bibnamefont
  {Bird}}, \bibinfo {author} {\bibfnamefont {C.~F.}\ \bibnamefont {Curtiss}},
  \bibinfo {author} {\bibfnamefont {R.~C.}\ \bibnamefont {Armstrong}},\ and\
  \bibinfo {author} {\bibfnamefont {O.}~\bibnamefont {Hassager}},\ }\href@noop
  {} {\emph {\bibinfo {title} {Dynamics of {P}olymeric {L}iquids. {V}olume 2:
  Kinetic theory}}},\ \bibinfo {edition} {2nd}\ ed.\ (\bibinfo  {publisher}
  {John Wiley and Sons, New York},\ \bibinfo {year} {1987})\BibitemShut
  {NoStop}%
\bibitem [{\citenamefont {Larson}(1988)}]{larson1988constitutive}%
  \BibitemOpen
  \bibfield  {author} {\bibinfo {author} {\bibfnamefont {R.~G.}\ \bibnamefont
  {Larson}},\ }\href@noop {} {\emph {\bibinfo {title} {Constitutive {E}quations
  for {P}olymer {M}elts and {S}olutions}}}\ (\bibinfo  {publisher}
  {Butterworths, Boston},\ \bibinfo {year} {1988})\BibitemShut {NoStop}%
\bibitem [{\citenamefont {Oldroyd}(1950)}]{oldroyd1950formulation}%
  \BibitemOpen
  \bibfield  {author} {\bibinfo {author} {\bibfnamefont {J.~G.}\ \bibnamefont
  {Oldroyd}},\ }\bibfield  {title} {\bibinfo {title} {On the formulation of
  rheological equations of state},\ }\href@noop {} {\bibfield  {journal}
  {\bibinfo  {journal} {Proc. R. Soc. A}\ }\textbf {\bibinfo {volume} {200}},\
  \bibinfo {pages} {523} (\bibinfo {year} {1950})}\BibitemShut {NoStop}%
\bibitem [{\citenamefont {Kuhn}(1934)}]{kuhn1934gestalt}%
  \BibitemOpen
  \bibfield  {author} {\bibinfo {author} {\bibfnamefont {W.}~\bibnamefont
  {Kuhn}},\ }\bibfield  {title} {\bibinfo {title} {{\"U}ber die gestalt
  fadenf{\"o}rmiger molek{\"u}le in l{\"o}sungen},\ }\href@noop {} {\bibfield
  {journal} {\bibinfo  {journal} {Kolloid-Z.}\ }\textbf {\bibinfo {volume}
  {68}},\ \bibinfo {pages} {2} (\bibinfo {year} {1934})}\BibitemShut {NoStop}%
\bibitem [{\citenamefont {Morozov}\ and\ \citenamefont
  {Spagnolie}(2015)}]{Intro_C_F}%
  \BibitemOpen
  \bibfield  {author} {\bibinfo {author} {\bibfnamefont {A.}~\bibnamefont
  {Morozov}}\ and\ \bibinfo {author} {\bibfnamefont {S.~E.}\ \bibnamefont
  {Spagnolie}},\ }\bibfield  {title} {\bibinfo {title} {Introduction to complex
  fluids},\ }in\ \href@noop {} {\emph {\bibinfo {booktitle} {Complex Fluids in
  Biological Systems}}},\ \bibinfo {editor} {edited by\ \bibinfo {editor}
  {\bibfnamefont {S.~E.}\ \bibnamefont {Spagnolie}}}\ (\bibinfo  {publisher}
  {Springer},\ \bibinfo {year} {2015})\ pp.\ \bibinfo {pages}
  {3--52}\BibitemShut {NoStop}%
\bibitem [{\citenamefont {Lumley}(1971)}]{lumley1971applicability}%
  \BibitemOpen
  \bibfield  {author} {\bibinfo {author} {\bibfnamefont {J.~L.}\ \bibnamefont
  {Lumley}},\ }\bibfield  {title} {\bibinfo {title} {Applicability of the
  {O}ldroyd constitutive equation to flow of dilute polymer solutions},\
  }\href@noop {} {\bibfield  {journal} {\bibinfo  {journal} {Phys. Fluids}\
  }\textbf {\bibinfo {volume} {14}},\ \bibinfo {pages} {2282} (\bibinfo {year}
  {1971})}\BibitemShut {NoStop}%
\bibitem [{\citenamefont {Hinch}\ and\ \citenamefont
  {Harlen}(2021)}]{hinch2021oldroyd}%
  \BibitemOpen
  \bibfield  {author} {\bibinfo {author} {\bibfnamefont {J.}~\bibnamefont
  {Hinch}}\ and\ \bibinfo {author} {\bibfnamefont {O.}~\bibnamefont {Harlen}},\
  }\bibfield  {title} {\bibinfo {title} {Oldroyd {B}, and not {A}?},\
  }\href@noop {} {\bibfield  {journal} {\bibinfo  {journal} {J. Non-Newtonian
  Fluid Mech.}\ }\textbf {\bibinfo {volume} {298}},\ \bibinfo {pages} {104668}
  (\bibinfo {year} {2021})}\BibitemShut {NoStop}%
\bibitem [{\citenamefont {Beris}(2021)}]{beris2021continuum}%
  \BibitemOpen
  \bibfield  {author} {\bibinfo {author} {\bibfnamefont {A.~N.}\ \bibnamefont
  {Beris}},\ }\bibfield  {title} {\bibinfo {title} {Continuum mechanics
  modeling of complex fluid systems following {O}ldroyd's seminal 1950 work},\
  }\href@noop {} {\bibfield  {journal} {\bibinfo  {journal} {J. Non-Newtonian
  Fluid Mech.}\ }\textbf {\bibinfo {volume} {298}},\ \bibinfo {pages} {104677}
  (\bibinfo {year} {2021})}\BibitemShut {NoStop}%
\bibitem [{\citenamefont {Datta}\ \emph {et~al.}(2022)\citenamefont {Datta},
  \citenamefont {Ardekani}, \citenamefont {Arratia}, \citenamefont {Beris},
  \citenamefont {Bischofberger}, \citenamefont {McKinley}, \citenamefont
  {Eggers}, \citenamefont {L\'opez-Aguilar}, \citenamefont {Fielding},
  \citenamefont {Frishman}, \citenamefont {Graham}, \citenamefont {Guasto},
  \citenamefont {Haward}, \citenamefont {Shen}, \citenamefont {Hormozi},
  \citenamefont {Morozov}, \citenamefont {Poole}, \citenamefont {Shankar},
  \citenamefont {Shaqfeh}, \citenamefont {Stark}, \citenamefont {Steinberg},
  \citenamefont {Subramanian},\ and\ \citenamefont
  {Stone}}]{datta2021perspectives}%
  \BibitemOpen
  \bibfield  {author} {\bibinfo {author} {\bibfnamefont {S.~S.}\ \bibnamefont
  {Datta}}, \bibinfo {author} {\bibfnamefont {A.~M.}\ \bibnamefont {Ardekani}},
  \bibinfo {author} {\bibfnamefont {P.~E.}\ \bibnamefont {Arratia}}, \bibinfo
  {author} {\bibfnamefont {A.~N.}\ \bibnamefont {Beris}}, \bibinfo {author}
  {\bibfnamefont {I.}~\bibnamefont {Bischofberger}}, \bibinfo {author}
  {\bibfnamefont {G.~H.}\ \bibnamefont {McKinley}}, \bibinfo {author}
  {\bibfnamefont {J.~G.}\ \bibnamefont {Eggers}}, \bibinfo {author}
  {\bibfnamefont {J.~E.}\ \bibnamefont {L\'opez-Aguilar}}, \bibinfo {author}
  {\bibfnamefont {S.~M.}\ \bibnamefont {Fielding}}, \bibinfo {author}
  {\bibfnamefont {A.}~\bibnamefont {Frishman}}, \bibinfo {author}
  {\bibfnamefont {M.~D.}\ \bibnamefont {Graham}}, \bibinfo {author}
  {\bibfnamefont {J.~S.}\ \bibnamefont {Guasto}}, \bibinfo {author}
  {\bibfnamefont {S.~J.}\ \bibnamefont {Haward}}, \bibinfo {author}
  {\bibfnamefont {A.~Q.}\ \bibnamefont {Shen}}, \bibinfo {author}
  {\bibfnamefont {S.}~\bibnamefont {Hormozi}}, \bibinfo {author} {\bibfnamefont
  {A.}~\bibnamefont {Morozov}}, \bibinfo {author} {\bibfnamefont {R.~J.}\
  \bibnamefont {Poole}}, \bibinfo {author} {\bibfnamefont {V.}~\bibnamefont
  {Shankar}}, \bibinfo {author} {\bibfnamefont {E.~S.~G.}\ \bibnamefont
  {Shaqfeh}}, \bibinfo {author} {\bibfnamefont {H.}~\bibnamefont {Stark}},
  \bibinfo {author} {\bibfnamefont {V.}~\bibnamefont {Steinberg}}, \bibinfo
  {author} {\bibfnamefont {G.}~\bibnamefont {Subramanian}},\ and\ \bibinfo
  {author} {\bibfnamefont {H.~A.}\ \bibnamefont {Stone}},\ }\bibfield  {title}
  {\bibinfo {title} {Perspectives on viscoelastic flow instabilities and
  elastic turbulence},\ }\href@noop {} {\bibfield  {journal} {\bibinfo
  {journal} {Phys. Rev. Fluids}\ }\textbf {\bibinfo {volume} {7}},\ \bibinfo
  {pages} {080701} (\bibinfo {year} {2022})}\BibitemShut {NoStop}%
\bibitem [{\citenamefont {Edwards}\ and\ \citenamefont
  {Beris}(2023)}]{edwards2023oldroyd}%
  \BibitemOpen
  \bibfield  {author} {\bibinfo {author} {\bibfnamefont {B.~J.}\ \bibnamefont
  {Edwards}}\ and\ \bibinfo {author} {\bibfnamefont {A.~N.}\ \bibnamefont
  {Beris}},\ }\bibfield  {title} {\bibinfo {title} {Oldroyd’s convected
  derivatives derived via the variational action principle and their
  corresponding stress tensors},\ }\href@noop {} {\bibfield  {journal}
  {\bibinfo  {journal} {J. Non-Newtonian Fluid Mech.}\ }\textbf {\bibinfo
  {volume} {316}},\ \bibinfo {pages} {105035} (\bibinfo {year}
  {2023})}\BibitemShut {NoStop}%
\bibitem [{\citenamefont {Hinch}(1974)}]{hinch1974mechanical}%
  \BibitemOpen
  \bibfield  {author} {\bibinfo {author} {\bibfnamefont {E.~J.}\ \bibnamefont
  {Hinch}},\ }\bibfield  {title} {\bibinfo {title} {Mechanical models of dilute
  polymer solutions for strong flows with large polymer deformations},\
  }\href@noop {} {\bibfield  {journal} {\bibinfo  {journal} {Colloques
  Internationaux du CNRS}\ }\textbf {\bibinfo {volume} {233}},\ \bibinfo
  {pages} {241} (\bibinfo {year} {1974})}\BibitemShut {NoStop}%
\bibitem [{\citenamefont {Hinch}(1977)}]{hinch1977mechanical}%
  \BibitemOpen
  \bibfield  {author} {\bibinfo {author} {\bibfnamefont {E.~J.}\ \bibnamefont
  {Hinch}},\ }\bibfield  {title} {\bibinfo {title} {Mechanical models of dilute
  polymer solutions in strong flows},\ }\href@noop {} {\bibfield  {journal}
  {\bibinfo  {journal} {Phys. Fluids}\ }\textbf {\bibinfo {volume} {20}},\
  \bibinfo {pages} {S22} (\bibinfo {year} {1977})}\BibitemShut {NoStop}%
\bibitem [{\citenamefont {de~Gennes}(1974)}]{de1974coil}%
  \BibitemOpen
  \bibfield  {author} {\bibinfo {author} {\bibfnamefont {P.~G.}\ \bibnamefont
  {de~Gennes}},\ }\bibfield  {title} {\bibinfo {title} {Coil-stretch transition
  of dilute flexible polymers under ultrahigh velocity gradients},\ }\href@noop
  {} {\bibfield  {journal} {\bibinfo  {journal} {J. Chem. Phys.}\ }\textbf
  {\bibinfo {volume} {60}},\ \bibinfo {pages} {5030} (\bibinfo {year}
  {1974})}\BibitemShut {NoStop}%
\bibitem [{\citenamefont {Bird}\ \emph {et~al.}(1980)\citenamefont {Bird},
  \citenamefont {Dotson},\ and\ \citenamefont {Johnson}}]{bird1980polymer}%
  \BibitemOpen
  \bibfield  {author} {\bibinfo {author} {\bibfnamefont {R.~B.}\ \bibnamefont
  {Bird}}, \bibinfo {author} {\bibfnamefont {P.~J.}\ \bibnamefont {Dotson}},\
  and\ \bibinfo {author} {\bibfnamefont {N.~L.}\ \bibnamefont {Johnson}},\
  }\bibfield  {title} {\bibinfo {title} {Polymer solution rheology based on a
  finitely extensible bead—spring chain model},\ }\href@noop {} {\bibfield
  {journal} {\bibinfo  {journal} {J. Non-Newtonian Fluid Mech.}\ }\textbf
  {\bibinfo {volume} {7}},\ \bibinfo {pages} {213} (\bibinfo {year}
  {1980})}\BibitemShut {NoStop}%
\bibitem [{\citenamefont {Fuller}\ and\ \citenamefont
  {Leal}(1980)}]{fuller1980flow}%
  \BibitemOpen
  \bibfield  {author} {\bibinfo {author} {\bibfnamefont {G.~G.}\ \bibnamefont
  {Fuller}}\ and\ \bibinfo {author} {\bibfnamefont {L.~G.}\ \bibnamefont
  {Leal}},\ }\bibfield  {title} {\bibinfo {title} {Flow birefringence of dilute
  polymer solutions in two-dimensional flows},\ }\href@noop {} {\bibfield
  {journal} {\bibinfo  {journal} {Rheol. Acta}\ }\textbf {\bibinfo {volume}
  {19}},\ \bibinfo {pages} {580} (\bibinfo {year} {1980})}\BibitemShut
  {NoStop}%
\bibitem [{\citenamefont {Fuller}\ and\ \citenamefont
  {Leal}(1981)}]{fuller1981effects}%
  \BibitemOpen
  \bibfield  {author} {\bibinfo {author} {\bibfnamefont {G.~G.}\ \bibnamefont
  {Fuller}}\ and\ \bibinfo {author} {\bibfnamefont {L.~G.}\ \bibnamefont
  {Leal}},\ }\bibfield  {title} {\bibinfo {title} {The effects of
  conformation-dependent friction and internal viscosity on the dynamics of the
  nonlinear dumbbell model for a dilute polymer solution},\ }\href@noop {}
  {\bibfield  {journal} {\bibinfo  {journal} {J. Non-Newtonian Fluid Mech.}\
  }\textbf {\bibinfo {volume} {8}},\ \bibinfo {pages} {271} (\bibinfo {year}
  {1981})}\BibitemShut {NoStop}%
\bibitem [{\citenamefont {Phan-Thien}\ \emph {et~al.}(1984)\citenamefont
  {Phan-Thien}, \citenamefont {Manero},\ and\ \citenamefont
  {Leal}}]{phan1984study}%
  \BibitemOpen
  \bibfield  {author} {\bibinfo {author} {\bibfnamefont {N.}~\bibnamefont
  {Phan-Thien}}, \bibinfo {author} {\bibfnamefont {O.}~\bibnamefont {Manero}},\
  and\ \bibinfo {author} {\bibfnamefont {L.~G.}\ \bibnamefont {Leal}},\
  }\bibfield  {title} {\bibinfo {title} {A study of conformation-dependent
  friction in a dumbbell model for dilute solutions},\ }\href@noop {}
  {\bibfield  {journal} {\bibinfo  {journal} {Rheol. Acta}\ }\textbf {\bibinfo
  {volume} {23}},\ \bibinfo {pages} {151} (\bibinfo {year} {1984})}\BibitemShut
  {NoStop}%
\bibitem [{\citenamefont {Dunlap}\ and\ \citenamefont
  {Leal}(1987)}]{dunlap1987dilute}%
  \BibitemOpen
  \bibfield  {author} {\bibinfo {author} {\bibfnamefont {P.~N.}\ \bibnamefont
  {Dunlap}}\ and\ \bibinfo {author} {\bibfnamefont {L.~G.}\ \bibnamefont
  {Leal}},\ }\bibfield  {title} {\bibinfo {title} {Dilute polystyrene solutions
  in extensional flows: {B}irefringence and flow modification},\ }\href@noop {}
  {\bibfield  {journal} {\bibinfo  {journal} {J. Non-Newtonian Fluid Mech.}\
  }\textbf {\bibinfo {volume} {23}},\ \bibinfo {pages} {5} (\bibinfo {year}
  {1987})}\BibitemShut {NoStop}%
\bibitem [{\citenamefont {Chilcott}\ and\ \citenamefont
  {Rallison}(1988)}]{chilcott1988creeping}%
  \BibitemOpen
  \bibfield  {author} {\bibinfo {author} {\bibfnamefont {M.~D.}\ \bibnamefont
  {Chilcott}}\ and\ \bibinfo {author} {\bibfnamefont {J.~M.}\ \bibnamefont
  {Rallison}},\ }\bibfield  {title} {\bibinfo {title} {Creeping flow of dilute
  polymer solutions past cylinders and spheres},\ }\href@noop {} {\bibfield
  {journal} {\bibinfo  {journal} {J. Non-Newtonian Fluid Mech.}\ }\textbf
  {\bibinfo {volume} {29}},\ \bibinfo {pages} {381} (\bibinfo {year}
  {1988})}\BibitemShut {NoStop}%
\bibitem [{\citenamefont {Harrison}\ \emph {et~al.}(1998)\citenamefont
  {Harrison}, \citenamefont {Remmelgas},\ and\ \citenamefont
  {Leal}}]{harrison1998dynamics}%
  \BibitemOpen
  \bibfield  {author} {\bibinfo {author} {\bibfnamefont {G.~M.}\ \bibnamefont
  {Harrison}}, \bibinfo {author} {\bibfnamefont {J.}~\bibnamefont
  {Remmelgas}},\ and\ \bibinfo {author} {\bibfnamefont {L.~G.}\ \bibnamefont
  {Leal}},\ }\bibfield  {title} {\bibinfo {title} {The dynamics of ultradilute
  polymer solutions in transient flow: {C}omparison of dumbbell-based theory
  and experiment},\ }\href@noop {} {\bibfield  {journal} {\bibinfo  {journal}
  {J. Rheol.}\ }\textbf {\bibinfo {volume} {42}},\ \bibinfo {pages} {1039}
  (\bibinfo {year} {1998})}\BibitemShut {NoStop}%
\bibitem [{\citenamefont {Remmelgas}\ \emph {et~al.}(1999)\citenamefont
  {Remmelgas}, \citenamefont {Singh},\ and\ \citenamefont
  {Leal}}]{remmelgas1999computational}%
  \BibitemOpen
  \bibfield  {author} {\bibinfo {author} {\bibfnamefont {J.}~\bibnamefont
  {Remmelgas}}, \bibinfo {author} {\bibfnamefont {P.}~\bibnamefont {Singh}},\
  and\ \bibinfo {author} {\bibfnamefont {L.~G.}\ \bibnamefont {Leal}},\
  }\bibfield  {title} {\bibinfo {title} {Computational studies of nonlinear
  elastic dumbbell models of {B}oger fluids in a cross-slot flow},\ }\href@noop
  {} {\bibfield  {journal} {\bibinfo  {journal} {J. Non-Newtonian Fluid Mech.}\
  }\textbf {\bibinfo {volume} {88}},\ \bibinfo {pages} {31} (\bibinfo {year}
  {1999})}\BibitemShut {NoStop}%
\bibitem [{\citenamefont {Venerus}\ and\ \citenamefont
  {\:Ottinger}(2018)}]{VenerusOttinger}%
  \BibitemOpen
  \bibfield  {author} {\bibinfo {author} {\bibfnamefont {D.}~\bibnamefont
  {Venerus}}\ and\ \bibinfo {author} {\bibfnamefont {H.}~\bibnamefont
  {\:Ottinger}},\ }\href@noop {} {\emph {\bibinfo {title} {A Modern Course in
  Transport Phenomena}}}\ (\bibinfo  {publisher} {Cambridge University Press,
  Cambridge},\ \bibinfo {year} {2018})\BibitemShut {NoStop}%
\bibitem [{\citenamefont {Batchelor}(2000)}]{batchelor2000introduction}%
  \BibitemOpen
  \bibfield  {author} {\bibinfo {author} {\bibfnamefont {G.~K.}\ \bibnamefont
  {Batchelor}},\ }\href@noop {} {\emph {\bibinfo {title} {An Introduction to
  Fluid Dynamics}}}\ (\bibinfo  {publisher} {Cambridge University Press},\
  \bibinfo {year} {2000})\BibitemShut {NoStop}%
\bibitem [{\citenamefont {Snoeijer}\ \emph {et~al.}(2020)\citenamefont
  {Snoeijer}, \citenamefont {Pandey}, \citenamefont {Herrada},\ and\
  \citenamefont {Eggers}}]{snoeijer2020relationship}%
  \BibitemOpen
  \bibfield  {author} {\bibinfo {author} {\bibfnamefont {J.~H.}\ \bibnamefont
  {Snoeijer}}, \bibinfo {author} {\bibfnamefont {A.}~\bibnamefont {Pandey}},
  \bibinfo {author} {\bibfnamefont {M.~A.}\ \bibnamefont {Herrada}},\ and\
  \bibinfo {author} {\bibfnamefont {J.}~\bibnamefont {Eggers}},\ }\bibfield
  {title} {\bibinfo {title} {The relationship between viscoelasticity and
  elasticity},\ }\href@noop {} {\bibfield  {journal} {\bibinfo  {journal}
  {Proc. R. Soc. Lond. A}\ }\textbf {\bibinfo {volume} {476}},\ \bibinfo
  {pages} {20200419} (\bibinfo {year} {2020})}\BibitemShut {NoStop}%
\bibitem [{\citenamefont {Brand}(1947)}]{LBrand}%
  \BibitemOpen
  \bibfield  {author} {\bibinfo {author} {\bibfnamefont {L.}~\bibnamefont
  {Brand}},\ }\href@noop {} {\emph {\bibinfo {title} {Vector and Tensor
  Analysis}}}\ (\bibinfo  {publisher} {John Wiley \& Sons, New York},\ \bibinfo
  {year} {1947})\BibitemShut {NoStop}%
\bibitem [{\citenamefont {Ericksen}(1960)}]{ericksen1960transversely}%
  \BibitemOpen
  \bibfield  {author} {\bibinfo {author} {\bibfnamefont {J.~L.}\ \bibnamefont
  {Ericksen}},\ }\bibfield  {title} {\bibinfo {title} {Transversely isotropic
  fluids},\ }\href@noop {} {\bibfield  {journal} {\bibinfo  {journal} {Kolloid
  Z.}\ }\textbf {\bibinfo {volume} {173}},\ \bibinfo {pages} {117} (\bibinfo
  {year} {1960})}\BibitemShut {NoStop}%
\bibitem [{\citenamefont {Gordon}\ and\ \citenamefont
  {E.}(1971)}]{gordon1971bead}%
  \BibitemOpen
  \bibfield  {author} {\bibinfo {author} {\bibfnamefont {R.~J.}\ \bibnamefont
  {Gordon}}\ and\ \bibinfo {author} {\bibfnamefont {E.~A.}\ \bibnamefont
  {E.}},\ }\bibfield  {title} {\bibinfo {title} {Bead-spring model of dilute
  polymer solutions: Continuum modifications and an explicit constitutive
  equation},\ }\href@noop {} {\bibfield  {journal} {\bibinfo  {journal} {J.
  Appl. Polym. Sci.}\ }\textbf {\bibinfo {volume} {15}},\ \bibinfo {pages}
  {1903} (\bibinfo {year} {1971})}\BibitemShut {NoStop}%
\bibitem [{\citenamefont {Gordon}\ and\ \citenamefont
  {Schowalter}(1972)}]{gordon1972anisotropic}%
  \BibitemOpen
  \bibfield  {author} {\bibinfo {author} {\bibfnamefont {R.~J.}\ \bibnamefont
  {Gordon}}\ and\ \bibinfo {author} {\bibfnamefont {W.~R.}\ \bibnamefont
  {Schowalter}},\ }\bibfield  {title} {\bibinfo {title} {Anisotropic fluid
  theory: a different approach to the dumbbell theory of dilute polymer
  solutions},\ }\href@noop {} {\bibfield  {journal} {\bibinfo  {journal}
  {Trans. Soc. Rheol.}\ }\textbf {\bibinfo {volume} {16}},\ \bibinfo {pages}
  {79} (\bibinfo {year} {1972})}\BibitemShut {NoStop}%
\bibitem [{\citenamefont {Jeffery}(1922)}]{jeffery1922motion}%
  \BibitemOpen
  \bibfield  {author} {\bibinfo {author} {\bibfnamefont {G.~B.}\ \bibnamefont
  {Jeffery}},\ }\bibfield  {title} {\bibinfo {title} {The motion of ellipsoidal
  particles immersed in a viscous fluid},\ }\href@noop {} {\bibfield  {journal}
  {\bibinfo  {journal} {Proc. R. Soc. A}\ }\textbf {\bibinfo {volume} {102}},\
  \bibinfo {pages} {161} (\bibinfo {year} {1922})}\BibitemShut {NoStop}%
\bibitem [{\citenamefont {Stone}(2017)}]{stone2017fundamentals}%
  \BibitemOpen
  \bibfield  {author} {\bibinfo {author} {\bibfnamefont {H.~A.}\ \bibnamefont
  {Stone}},\ }\bibfield  {title} {\bibinfo {title} {Fundamentals of fluid
  dynamics with an introduction to the importance of interfaces},\ }in\
  \href@noop {} {\emph {\bibinfo {booktitle} {Soft Interfaces, Lecture Notes of
  the Les Houches Summer School}}},\ \bibinfo {editor} {edited by\ \bibinfo
  {editor} {\bibfnamefont {L.}~\bibnamefont {Bocquet}}, \bibinfo {editor}
  {\bibfnamefont {D.}~\bibnamefont {Qu{\'e}r{\'e}}}, \bibinfo {editor}
  {\bibfnamefont {T.~A.}\ \bibnamefont {Witten}},\ and\ \bibinfo {editor}
  {\bibfnamefont {L.~F.}\ \bibnamefont {Cugliandolo}}}\ (\bibinfo  {publisher}
  {Oxford University Press, New York},\ \bibinfo {year} {2017})\ Chap.~\bibinfo
  {chapter} {1}, pp.\ \bibinfo {pages} {3--76}\BibitemShut {NoStop}%
\bibitem [{\citenamefont {Eggers}\ \emph {et~al.}(2023)\citenamefont {Eggers},
  \citenamefont {Liverpool},\ and\ \citenamefont
  {Mietke}}]{eggers2023rheology}%
  \BibitemOpen
  \bibfield  {author} {\bibinfo {author} {\bibfnamefont {J.}~\bibnamefont
  {Eggers}}, \bibinfo {author} {\bibfnamefont {T.~B.}\ \bibnamefont
  {Liverpool}},\ and\ \bibinfo {author} {\bibfnamefont {A.}~\bibnamefont
  {Mietke}},\ }\bibfield  {title} {\bibinfo {title} {Rheology of suspensions of
  flat elastic particles},\ }\href@noop {} {\bibfield  {journal} {\bibinfo
  {journal} {preprint arXiv:2304.02980}\ } (\bibinfo {year}
  {2023})}\BibitemShut {NoStop}%
\bibitem [{\citenamefont {Zaremba}(1903)}]{zaremba1903remarques}%
  \BibitemOpen
  \bibfield  {author} {\bibinfo {author} {\bibfnamefont {S.}~\bibnamefont
  {Zaremba}},\ }\bibfield  {title} {\bibinfo {title} {Remarques sur les travaux
  de {M.} {N}atanson relatifs {\`a} la th{\'e}orie de la viscosit{\'e}},\
  }\href@noop {} {\bibfield  {journal} {\bibinfo  {journal} {Bull. Int. Acad.
  Sci. Crac.}\ ,\ \bibinfo {pages} {85}} (\bibinfo {year} {1903})}\BibitemShut
  {NoStop}%
\bibitem [{\citenamefont {Jaumann}(1911)}]{jaumann1911geschlossenes}%
  \BibitemOpen
  \bibfield  {author} {\bibinfo {author} {\bibfnamefont {G.}~\bibnamefont
  {Jaumann}},\ }\bibfield  {title} {\bibinfo {title} {Geschlossenes system
  physicalisher und chemischer differentialgesetze},\ }\href@noop {} {\bibfield
   {journal} {\bibinfo  {journal} {Sitzber. Akad. Wiss. Wien (IIa)}\ }\textbf
  {\bibinfo {volume} {120}},\ \bibinfo {pages} {385} (\bibinfo {year}
  {1911})}\BibitemShut {NoStop}%
\bibitem [{\citenamefont {Rallison}\ and\ \citenamefont
  {Hinch}(1988)}]{rallison1988we}%
  \BibitemOpen
  \bibfield  {author} {\bibinfo {author} {\bibfnamefont {J.~M.}\ \bibnamefont
  {Rallison}}\ and\ \bibinfo {author} {\bibfnamefont {E.~J.}\ \bibnamefont
  {Hinch}},\ }\bibfield  {title} {\bibinfo {title} {Do we understand the
  physics in the constitutive equation?},\ }\href@noop {} {\bibfield  {journal}
  {\bibinfo  {journal} {J. Non-Newtonian Fluid Mech.}\ }\textbf {\bibinfo
  {volume} {29}},\ \bibinfo {pages} {37} (\bibinfo {year} {1988})}\BibitemShut
  {NoStop}%
\bibitem [{\citenamefont {Weady}\ \emph {et~al.}(2022)\citenamefont {Weady},
  \citenamefont {Stein},\ and\ \citenamefont
  {Shelley}}]{weady2022thermodynamically}%
  \BibitemOpen
  \bibfield  {author} {\bibinfo {author} {\bibfnamefont {S.}~\bibnamefont
  {Weady}}, \bibinfo {author} {\bibfnamefont {D.~B.}\ \bibnamefont {Stein}},\
  and\ \bibinfo {author} {\bibfnamefont {M.~J.}\ \bibnamefont {Shelley}},\
  }\bibfield  {title} {\bibinfo {title} {Thermodynamically consistent
  coarse-graining of polar active fluids},\ }\href@noop {} {\bibfield
  {journal} {\bibinfo  {journal} {Phys. Rev. Fluids}\ }\textbf {\bibinfo
  {volume} {7}},\ \bibinfo {pages} {063301} (\bibinfo {year}
  {2022})}\BibitemShut {NoStop}%
\bibitem [{\citenamefont {Gao}\ \emph {et~al.}(2017)\citenamefont {Gao},
  \citenamefont {Betterton}, \citenamefont {Jhang},\ and\ \citenamefont
  {Shelley}}]{gao2017analytical}%
  \BibitemOpen
  \bibfield  {author} {\bibinfo {author} {\bibfnamefont {T.}~\bibnamefont
  {Gao}}, \bibinfo {author} {\bibfnamefont {M.~D.}\ \bibnamefont {Betterton}},
  \bibinfo {author} {\bibfnamefont {A.-S.}\ \bibnamefont {Jhang}},\ and\
  \bibinfo {author} {\bibfnamefont {M.~J.}\ \bibnamefont {Shelley}},\
  }\bibfield  {title} {\bibinfo {title} {Analytical structure, dynamics, and
  coarse graining of a kinetic model of an active fluid},\ }\href@noop {}
  {\bibfield  {journal} {\bibinfo  {journal} {Phys. Rev. Fluids}\ }\textbf
  {\bibinfo {volume} {2}},\ \bibinfo {pages} {093302} (\bibinfo {year}
  {2017})}\BibitemShut {NoStop}%
\bibitem [{\citenamefont {Hohenegger}\ and\ \citenamefont
  {Shelley}(2011)}]{hohenegger2011dynamics}%
  \BibitemOpen
  \bibfield  {author} {\bibinfo {author} {\bibfnamefont {C.}~\bibnamefont
  {Hohenegger}}\ and\ \bibinfo {author} {\bibfnamefont {M.~J.}\ \bibnamefont
  {Shelley}},\ }\href@noop {} {\emph {\bibinfo {title} {Dynamics of Complex
  Bio-Fluids}}}\ (\bibinfo  {publisher} {Oxford University Press, Oxford},\
  \bibinfo {year} {2011})\BibitemShut {NoStop}%
\end{thebibliography}

\end{document}